# Neutron Data Evaluation of $^{243}$Am

V.M. Maslov[1], V.G. Pronyaev[2], N.A. Tetereva[1],
K.I. Zolotarev[2]

[1] Joint Institute of Power and Nuclear Research – SOSNY
220109, Minsk, Byelorussia (during 2008-2011)
[2] Institute of Physics and Power Engineering, 249020, Obninsk, Russia

## Abstract

The diverse measured data base of n+$^{243}$Am was evaluated using a statistical theory and genera-lized least squares codes. Consistent description of total, capture and fission measured data provides an important constraint for the inelastic scattering cross section. Important constraints for the measured capture cross section in the 0.15-300 keV energy range come from the average radiative and neutron $S_0$ and $S_1$ strength functions. The evaluated inelastic cross sections of available evaluations are in severe disagreement, predicted change of the inelastic cross section shape at $E_n$ ~1.5 MeV is attributed to the sharp increase of the level density of the residual odd-even nuclide $^{241}$Am due to the onset of three-quasi-particle excitations.

The influence of exclusive (n, xnf) pre-fission neutrons on prompt fission neutron spectra (PFNS) and (n, xn) spectra is modelled. Contributions of emissive/non-emissive fission and exclusive spectra of (n, xnf) reactions are defined by a consistent description of the $^{241}$Am(n, F), $^{241}$Am(n, 2n).

This work is performed under the project Agreement B-1604 during 2008-2011 with the International Science and Technology Center (Moscow). The financing party is EU. Partial support of International Atomic Energy Agency under Research Contract 14809 is acknowledged. Data file is at https://www-nds.iaea.org/minskact/data/original/za095243

## Contents

## 1. Introduction

Americium-243 ($t_{1/2}$ =7370 yr) evolves in Pu-containing nuclear fuels at increased burn-up, after β—decay of $^{241}$Pu ($t_{1/2}$ =14.4 yr) in a chain, starting from $^{241}$Am. The transmutation of the $^{241}$Am in thermal power reactors is accompanied by the neutron capture cross sections in the reaction chains $^{241}$Am(n, γ) $^{242m(g)}$Am. The neutron capture reaction $^{241}$Am(n, γ) populates either the 16-h ground state $^{242g}$Am or the $^{242m}$Am isomer with $t_{1/2}$ =141 yr, in both cases build-up of $^{243}$Am is possible. The former state $^{242g}$Am subsequently β—-decays to $^{242}$Cm. The capture reaction $^{241}$Am(n, γ)$^{242m}$Am influences the neutron activity of the spent fuel due to spontaneous fission of $^{242m}$Am. It gives a path for the $^{244}$Cm yield via $^{242m}$Am(n,γ)$^{243}$Am(n,γ)$^{244m}$Am(β—(ε))$^{244}$Cm($^{244}$Pu) or $^{242m}$Am(n,γ)$^{243}$Am(n,γ)$^{244g}$Am(β-) $^{244}$Cm. If not the forbidden β—-decay of $^{242m}$Am state, the major path of $^{244}$Cm build-up would not exist. The yield of the $^{244m(g)}$Am states influences both alpha- and neutron activity of the spent fuel [1, 2].

Repository or transmutation of $^{243}$Am as one of the constituent of the spent fuel needs rather diverse knowledge base. Namely, the $^{243}$Am neutron-induced fission, capture, inelastic scattering, (n,2n) cross sections, branching ratio for the yields of short-lived $^{244m}$Am and long-lived $^{244g}$Am states of $^{243}$Am(n, γ) reaction, branching ratio for the yields of short-lived $^{242g}$Am and long-lived $^{242m}$Am states of $^{243}$Am(n, 2n) reaction. Prompt fission neutron spectrum (PFNS) and prompt fission neutron multiplicity are another important items, measurements of the former for the $^{243}$Am(n, F) reaction are unavailable at the moment.

The improvements of the nuclear reaction modeling and nuclear parameter systematic, developed based on neutron data description of neutron data for major actinides $^{232}$Th, $^{233}$U, $^{235}$U, $^{238}$U, $^{239}$Pu and Z-odd target $^{239}$Np provide a sound basis for critical assessment of the (n, F), (n, γ), (n, n), (n, n') cross sections and secondary neutron spectra for the n+$^{243}$Am interaction. The main reasons of improvement might be consistent description of total, fission and capture data in 0.15 keV – 300 keV energy range. For neutron capture reaction on $^{243}$Am target nuclide in the unresolved resonance and fast neutron energy ranges the methods, proven in case of $^{232}$Th(n,γ) [3], $^{238}$U(n, γ) [4] and $^{237}$Np [5] data analysis would be used. Disentangling of the model deficiencies and model parameter uncertainties, when measured cross section data base fits are rather poor, especially when the data are scattering and there are systematic shifts between different data sets, turns out to be a major problem in case of Z-odd actinides like $^{237}$Np or $^{241,243}$Am. Important constraints for the calculated capture cross section come from the average radiation width and neutron strength functions $S_0$ and $S_1$. For the $^{243}$Am+n interaction we have used almost the same optical potential, which allowed consistent description of total, fission and partial inelastic scattering data in 1~3 MeV incident neutron energy range for $^{237}$Np target nuclide. It provided an important constraint for the absorption cross section, which is quite important for the robust estimate of the capture cross section in keV- energy range.

Consistent description of $^{241}$Am(n, 2n) and $^{241}$Am (n, F) cross sections was considered an important constraint in view of large scatter and systematic shifts of fission data. $^{243}$Am (n, F) observed fission cross section data are scattered even more, calculated cross section is represented as a superposition of the (n, f) and (n, xnf) reactions, with simultaneous calculation of exclusive neutron spectra of (n, xn) and (n, xnf) reactions. That approach provided a robust estimates of prompt fission neutron spectra and their average energies for major actinides $^{232}$Th, $^{233}$U, $^{235}$U, $^{238}$U, $^{239}$Pu [6, 7] and Z-odd target $^{237}$Np [5, 7]. The average energies of prompt fission neutron spectra (PFNS) and PFNS itself [5, 7] for $^{237}$Np(n, F) reaction are consistent with measured data base. That is a strong impetus to provide a new evaluation of PFNS for $^{243}$Am (n, F).

Realistic assessment of the uncertainties of $^{243}$Am evaluated data should take into account the results of the consistent description of total, fission, capture and inelastic scattering cross sections with nuclear reaction theory. Purpose of the present evaluation is to clear out whether the available data on total and partial cross sections and average resonance parameters could be described (reproduced) consistently.

## 2. Resonance Parameters

Here we will briefly review the status of resolved neutron resonance parameters of $^{243}$Am. Resolved and unresolved resonance parameters are adopted from our previous evaluation [8], i.e., resolved resonance parameters for multilevel Breit-Wigner formalism (up to 150 eV) are adopted. Thermal total, capture and fission cross sections and resonance integrals are shown in Table 2.1.

**Table 2.1** Thermal total, elastic, capture and fission cross sections and resonance integrals

| Reaction | $\sigma^{th}$,barn | RI | $\sigma^{th}$,barn | RI | $\sigma^{th}$,barn | RI |
|---|---|---|---|---|---|---|
| | Present | | JENDL-4.0 | | ENDF/B-VII.0 | |
| Total | 84.232 | | 85.831 | | 83.6875 | |
| Elastic | 7.4642 | | 6.49 | | 8.53775 | |
| Fission | .00638 | 7.4654 | .00816 | 6.31 | .007389 | 8.63501 |
| Capture | 76.704 | 1781.18 | 79.259 | 2040 | 75.0759 | 1814.44 |

## 3. Evaluation of neutron capture and fission cross sections for $^{243}$Am with generalized least-squares method

The approach developed by W. Poenitz, the GMA code and database of experimental results were used in the last evaluation of the neutron cross section standards [9, 10]. The most measurements of the neutron capture and fission cross sections at $^{243}$Am have been done relative these standard reactions. These standards can be used in new evaluation of $^{243}$Am(n, γ) and $^{243}$Am(n, f) reaction cross sections in two ways. First, the results of standards evaluation (cross sections and covariance matrices of their uncertainties) may be used as a prior in a Bayesian approach. Uninformative prior should be assigned to $^{243}$Am(n, γ) and $^{243}$Am(n, f) cross sections. Using Bayesian procedure a posterior evaluation for combined standards and $^{243}$Am cross sections can be obtained by the least-squares fit adding the experimental data. In the second method, the generalized least-squares fit can be done for combined standard and $^{243}$Am(n, γ) and $^{243}$Am(n, f) cross sections. The final evaluation will contain the standards, which probably will be slightly modified, and evaluated $^{243}$Am(n, γ) and $^{243}$Am(n, f) cross sections with covariance matrix of their uncertainties, including the blocks with all cross-reaction correlations. The results obtained in both approaches should be almost coincident.

The combined fit with only standard for $^{197}$Au(n, γ) was used in the evaluation of capture cross section $^{243}$Am(n, γ). Because the $^{243}$Am(n, γ) experimental data include only the results of shape cross section measurements and absolute ratio of $^{243}$Am(n, γ) to $^{197}$Au(n, γ) cross sections, this is a rather good approximation. This is endorsed by the result of such evaluation when changes in the values of the $^{197}$Au(n, γ) reaction standard due to influence of the $^{243}$Am(n, γ) data are within 0.1 − 0.2 %.

## 3.1 $^{243}$Am capture cross section evaluation

The experimental data used in the evaluation are retrieved from EXFOR database and given in Table 1. Because the measurements of thermal capture cross section were de-coupled from measurements at other energies and include mostly Maxwellian neutron spectrum averaged cross section with kT=0.0253 eV, the evaluation at thermal energy point 0.0253 eV were done separately. The energy measurements were done by two groups, by L.W. Weston and J.H. Todd [11], which because of their normalization can be used only as shape type of data (see comment in table 1). Data by K. Wisshak and F. Kaeppeler [12], which presents detailed studies with variable parameters of measurement (16 data sets), but because of this, their systematical uncertainties are strongly correlated.

Table 3.1. The datasets for $^{243}$Am(n,γ) reaction cross section included in the GMA combined fit with $^{197}$Au(n,γ) reaction standard.

| Data set numb | First author | | Data | Type of measurem. | Energy range covered, MeV | Comments |
|---|---|---|---|---|---|---|
| 6001 | L.W. Weston | 22951002 (1985) | $^{243}$Am(n,γ) | shape | 0.0139 – 0.0838 | Data were used as a shape type of data, because their shape is well determined with $^{10}$B(n,α) neutron flux monitor which has well determined 1/v energy dependence in the energy range of the measurements. But final normalization was done on σ·sqrt(E$_n$) value in the energy range 0.02 – 0.1 eV for $^{243}$Am(n,γ) from ENDF/B-V library. |
| 6002 | K. Wisshak | 21863002 (1983) | $^{243}$Am(n,γ)/ $^{197}$Au(n, γ) | absolute | 0.0057 – 0.092 | Systematical uncertainties (LERC and MERC) is correlated with those of other measurements compiled in this entry because of similar conditions of measurements. |
| 6003 | K. Wisshak | 21863002 (1983) | $^{243}$Am(n,γ)/ $^{197}$Au(n, γ) | absolute | 0.0057 – 0.092 | Systematical uncertainties (LERC and MERC) is correlated with those of other measurements compiled in this entry because of similar conditions of measurements. |
| 6004 | K. Wisshak | 21863003 (1983) | $^{243}$Am(n,γ)/ $^{197}$Au(n, γ) | absolute | 0.0057 – 0.094 | Systematical uncertainties (LERC and MERC) is correlated with those of other measurements compiled in this entry because of similar conditions of measurements. |
| 6005 | K. Wisshak | 21863003 (1983) | $^{243}$Am(n,γ)/ $^{197}$Au(n, γ) | absolute | 0.0057 – 0.094 | Systematical uncertainties (LERC and MERC) is correlated with those of other measurements compiled in this entry because of similar conditions of measurements. |
| 6006 | K. Wisshak | 21863004 (1983) | $^{243}$Am(n,γ)/ $^{197}$Au(n, γ) | absolute | 0.0056 – 0.092 | Was not used in the fit because of discrepancy with other data of these authors |
| 6007 | K. Wisshak | 21863004 (1983) | $^{243}$Am(n,γ)/ $^{197}$Au(n, γ) | absolute | 0.0056 – 0.092 | Systematical uncertainties (LERC and MERC) is correlated with |
| | | | | | | those of other measurements compiled in this entry because of similar conditions of measurements. |
| 6008 | K. Wisshak | 21863005 (1983) | $^{243}$Am(n,$\gamma$)/ $^{197}$Au(n, $\gamma$) | absolute | $0.0058 - 0.092$ | Systematical uncertainties (LERC and MERC) is correlated with those of other measurements compiled in this entry because of similar conditions of measurements. |
| 6009 | K. Wisshak | 21863005 (1983) | $^{243}$Am(n,$\gamma$)/ $^{197}$Au(n, $\gamma$) | absolute | $0.0058 - 0.092$ | Systematical uncertainties (LERC and MERC) is correlated with those of other measurements compiled in this entry because of similar conditions of measurements. |
| 6010 | K. Wisshak | 21863006 (1983) | $^{243}$Am(n,$\gamma$)/ $^{197}$Au(n, $\gamma$) | absolute | $0.007 - 0.093$ | Systematical uncertainties (LERC and MERC) is correlated with those of other measurements compiled in this entry because of similar conditions of measurements. |
| 6011 | K. Wisshak | 21863006 (1983) | $^{243}$Am(n,$\gamma$)/ $^{197}$Au(n, $\gamma$) | absolute | $0.007 - 0.093$ | Systematical uncertainties (LERC and MERC) is correlated with those of other measurements compiled in this entry because of similar conditions of measurements. |
| 6012 | K. Wisshak | 21863007 (1983) | $^{243}$Am(n,$\gamma$)/ $^{197}$Au(n, $\gamma$) | absolute | $0.007 - 0.095$ | Systematical uncertainties (LERC and MERC) is correlated with those of other measurements compiled in this entry because of similar conditions of measurements. |
| 6013 | K. Wisshak | 21863007 (1983) | $^{243}$Am(n,$\gamma$)/ $^{197}$Au(n,g) | absolute | $0.007 - 0.095$ | Systematical uncertainties (LERC and MERC) is correlated with those of other measurements compiled in this entry because of similar conditions of measurements. |
| 6014 | K. Wisshak | 21863008 (1983) | $^{243}$Am(n,$\gamma$)/ $^{197}$Au(n, $\gamma$) | absolute | $0.034 - 0.226$ | Systematical uncertainties (LERC and MERC) is correlated with those of other measurements compiled in this entry because of similar conditions of measurements. |
| 6015 | K. Wisshak | 21863008 (1983) | $^{243}$Am(n,$\gamma$)/ $^{197}$Au(n, $\gamma$) | absolute | $0.034 - 0.226$ | Systematical uncertainties (LERC and MERC) is correlated with those of other measurements compiled in this entry because of similar conditions of measurements. |
| 6016 | K. Wisshak | 21863009 (1983) | $^{243}$Am(n,$\gamma$)/ $^{197}$Au(n, $\gamma$) | absolute | $0.038 - 0.218$ | Systematical uncertainties (LERC and MERC) is correlated with those of other measurements compiled in this entry because of similar conditions of measurements. |

| 6017 | K. Wisshak | 21863009 (1983) | $^{243}$Am(n,γ)/ $^{197}$Au(n, γ) | absolute | 0.038 − 0.218 | Systematical uncertainties (LERC and MERC) is correlated with those of other measurements compiled in this entry because of similar conditions of measurements. |
|------|-----------|-----------------|-----------------------------------|----------|---------------|--------------------------------------------------------------------------------------------------------------------------------------------------------------------|

The result of the evaluation are shown at Fig. 3.1 and 3.2 together with experimental data and latest evaluation of $^{243}$Am capture cross section done for JENDL-4.0 [13] library. Data on Fig. 3.1 are shown in the log-log scale. Weston and Todd [11] shape type data taken at their experimental energy nodes tuned with the normalization coefficient obtained in the fit with GMA. Wisshak et al. [12] data are converted from ratio to $^{197}$Au(n, γ) using $^{197}$Au(n, γ) cross section obtained in this combined fit, which, as it was mentioned before, differs from previously evaluated standard only in few points at 0.1 − 0.2 % level. The small uncertainty of $^{243}$Am(n, γ) cross section is at the level of 5 − 6 % below 10 keV, then reduces to 2 − 3 % in the energy range 10 − 80 keV, and increases again to 4 − 6 % in 80 − 230 keV energy range. Coefficients of the correlation matrix of the evaluated uncertainties are varied between 0.2 − 0.6 for range where Weston et al. [11] data determine the evaluation and are low for part of the matrix describing correlations between uncertainties of the data above and below 80 keV. General chi-square per degree of freedom in the fit is at the level of 0.9. Data at Fig 3.2 are shown as cross section multiplied by square root from the incident neutron energy, to remove close to "$1/v$" energy dependence observed at these energies.

Evaluation of the same experimental data were done independently (including analysis of experimental data and their uncertainties) using PADE analytical model approximation. In this evaluation Weston et al. [11] data were considered as shape type of data; this explains the difference in the low-energy part of the cross section. The results of Weston et al. [12] measurements show some fluctuations, which may have physical origine because of the low level density for capture channel in this energy region. It is interesting to mention, that JENDL-4.0 evaluation in the energy range between 80 and 200 keV is about 10% higher GMA and PADE evaluations.

### 3.2 $^{243}$Am fission cross section evaluation

Data from JENDL-4.0 [13] library for $^{243}$Am(n,f) reaction were used as a prior for ofr experimental data [14-27] to the energy nodes in the GMA fit. The experimental data selected for the fit given in the Table 3.2. Some data presented by the authors as results of absolute measurements (data set 7020 and 7023) were used as shape data, because of obvious problems with the normalization. The results of the GMA fit in comparison with JENDL-4.0 [13] evaluation and experimental data are shown in Figs. 3.3 − 3.7. When uncertainties assigned by authors are used, the chi-squared value per degree of freedom obtained in the fit for $^{243}$Am(n, f) is about 4, with the chi-squared value per degree of freedom for all data in the fit at the level of 1. The outlaying data were determined as data deviating from the result of the fit at more that 3σ -level in separate points and more than 2σ in few sequential energy points. The additional medium energy range correlation components of the uncertainties were added to the covariance matrices of the uncertainties of these data. As a result, the final chi-squared value per degree of freedom obtained in the fit of the data sets for $^{243}$Am(n, f) cros section was of the order of 1.

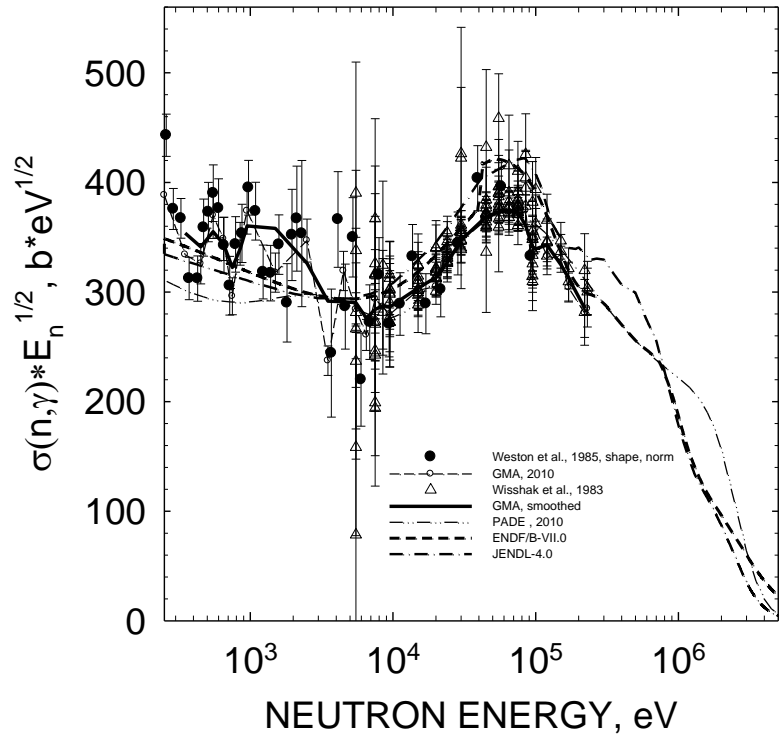

Fig. 3.1. Comparison of the results of the GMA evaluation with experimental data and JENDL-4.0 and PADE evaluation. GMA evaluation was smoothed using simple 3-point smoothing scheme.

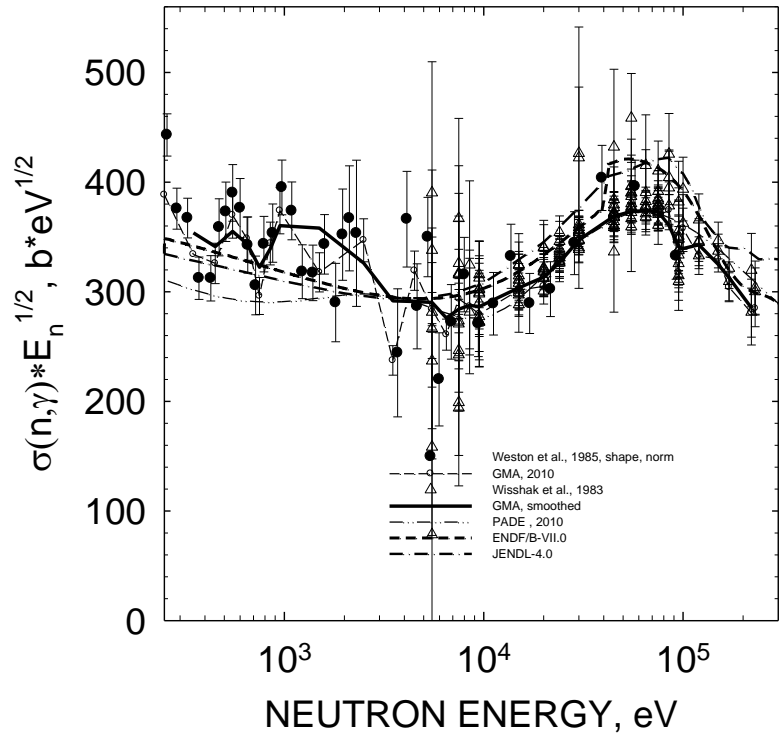

Fig. 3.2. Comparison of the results of the GMA evaluation with experimental data and JENDL-4.0 and PADE evaluations. Non-smoothed GMA evaluation shown by dashed curve.

Table 3.2. The data sets for $^{243}$Am(n,f) reaction cross section included in the GMA combined fit.

| Data set numb | First author | EXFOR entry, date | Data | Type of measurem. | Energy range covered, MeV | Comments |
|---|---|---|---|---|---|---|
| 7001 | P.A. Seeger | 10063 (1970) | $^{243}$Am(n,f) | Absolute | 49 eV – 30 MeV | Only data above 0.1 MeV are used in GMA fit because of large uncertainty and background below 0.1 MeV. 235U(n,f) was used for flux normalization. Because systematical uncertainty is substantially larger than the uncertainty of the standard data were not converted to $^{243}$Am(n,f)/$^{235}$U(n,f) ratio |
| 7002 | J.W. Behrens | 10652 (1981) | $^{243}$Am(n,f)/ $^{235}$U(n,f) | Shape of ratio | 0.2 – 30 MeV | Shape of ratio type of data was used because authors normalized data through ratio of cross sections in a wide energy interval (1.75 – 4.0 MeV). Shape of ratio was a primarely measured quantity |
| 7003 | D.K. Butler | 12543 (1961) | $^{243}$Am(n,f) | Shape | 0.03 – 1.67 MeV | Shape data with minimal error assigned 5 % . No nunerical data about $^{235}$U(n,f) monitor used. |
| 7004 | K.Wisshak | 21863013 (1983) | $^{243}$Am(n,f)/ $^{235}$U(n,f) | Absolute ratio | 0.0125 – 0.08 MeV | Data from runs I – IV, correlated with data sets 7005, 7006, 7007 |
| 7005 | K.Wisshak | 21863014 (1983) | $^{243}$Am(n,f)/ $^{235}$U(n,f) | Absolute ratio | 0.0125 – 0.08 MeV | Data from runs V and VI, correlated with data sets 7004, 7006, 7007 |
| 7006 | K.Wisshak | 21863015 (1983) | $^{243}$Am(n,f)/ $^{235}$U(n,f) | Absolute ratio | 0.035 – 0.2 MeV | Data from runs V and VI, correlated with data sets 7004, 7005, 7007 |
| 7007 | K.Wisshak | 21863016 (1983) | $^{243}$Am(n,f)/ $^{235}$U(n,f) | Absolute ratio | 0.035 – 0.225 MeV | Data from runs V and VI, correlated with data sets 7004, 7005, 7006 |
| 7008 | H.Terayama | 22024 (1986) | $^{243}$Am(n,f)/ $^{235}$U(n,f) | Absolute ratio | 4.31 – 6.83 MeV | d-d neutron source. Correlated with data sets 7009 |
| 7009 | H.Terayama | 22024 (1986) | $^{243}$Am(n,f)/ $^{235}$U(n,f) | Absolute ratio | 1.06 – 2.88 MeV | p-T neutron source. Correlated with data sets 7009 |
| 7010 | H. Knitter | 22032 (1988) | $^{243}$Am(n,f) | Absolute | 3.76 – 9.92 MeV | d-d neutron source. Taken as absolute because of large uncertainty comparing with $^{235}$U(n,f) standard changes in this energy range. Correlated with data set 7011 and 7012. |
| 7011 | H. Knitter | 22032 (1988) | $^{243}$Am(n,f) | Absolute | 0.897 – 3.96 MeV | p-T neutron source. Taken as absolute because of large uncertainty comparing with $^{235}$U(n,f) standard changes in this energy range. Correlated with data set 7010 and 7012. |
| 7012 | H. Knitter | 22032 (1988) | $^{243}$Am(n,f) | Absolute | 0.335 – 1.132 MeV | p-$^7$Li neutron source. Taken as absolute because of large uncertainty comparing with $^{235}$U(n,f) standard changes in this |
| | | | | | | energy range. Correlated with data set 7010 and 7011. |
| 7013 | F. Manabe | 22282 (1988) | $^{243}$Am(n,f)/ $^{235}$U(n,f) | Absolute ratio | 14 – 14.9 MeV | Correlated data. |
| 7014 | M. Aiche | 22993 (2007) | $^{243}$Am(n,f) | Absolute | 4.77 – 7.35 MeV | d-d neutrons. Taken as absolute because of large uncertainty comparing with $^{235}$U(n,f) standard changes in this energy range. Correlated with data set 7015. |
| 7015 | M. Aiche | 22993 (2007) | $^{243}$Am(n,f) | Absolute | 4.77 – 7.35 MeV | p-t neutrons. Taken as absolute because of large uncertainty comparing with $^{235}$U(n,f) standard changes in this energy range. Correlated with data set 7014. |
| 7016 | E.F. Fomushkin | 40779 (1967) | $^{243}$Am(n,f) | Absolute | 14.5 MeV | Renormalized to the standard values for 238U(n,f). Correlated with data set 7017. |
| 7017 | E.F. Fomushkin | 40779 (1967) | $^{243}$Am(n,f) | Absolute | 14.5 MeV | Renormalized to the standard values for 237Np(n,f) and 238U(n,f). Correlated with data set 7016. |
| 7018 | B.I. Fursov | 40837 (1985) | $^{243}$Am(n,f)/ $^{239}$Pu(n,f) | Absolute ratio | 0.135 – 7.40 MeV | Statistical uncertainty was estmated from known total and syatematical |
| 7019 | E.F. Fomushkin | 40856 (1984) | $^{243}$Am(n,f)/ $^{235}$U(n,f) | Absolute ratio | 14.8 MeV | |
| 7020 | A.A. Goverdovskiy | 40912 (1986) | $^{243}$Am(n,f)/ $^{235}$U(n,f) | Shape of ratio | 4.97 – 10.41 MeV | Shape of ratio was used because the normalization of ratio was not discussed |
| 7021 | V.Ya. Golovnya | 41361 (1994) | $^{243}$Am(n,f) | Absolute | 14.7 MeV | Associated particles method |
| 7022 | A.V. Fomichev | 41444 (2004) | $^{243}$Am(n,f)/ $^{239}$U(n,f) | Shape of ratio | 0.6 – 360 MeV | |
| 7023 | A.B. Laptev | 41487 (2007) | $^{243}$Am(n,f)/ $^{235}$U(n,f) | Shape of ratio | 0.58 – 196.2 MeV | The data were used as a shape of ratio data because of the problem with ratio normalization |

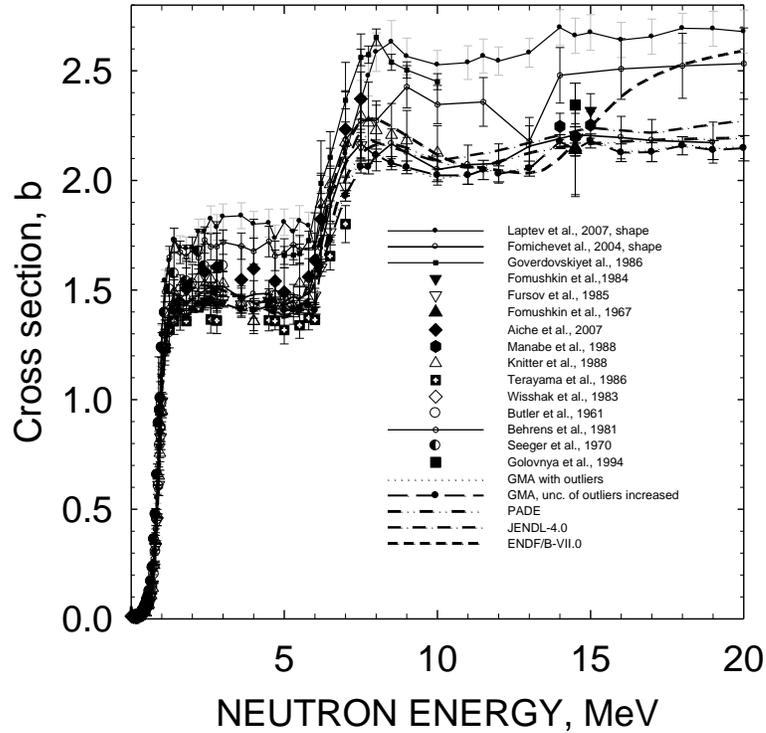

Fig. 3.3. Comparison of the results of the GMA evaluation with JENDL-4.0 [13] and experimental data. Experimental data shown as symbols connected by lines are non-normalized shape type of data. X4 is a number of EXFOR entry (or sub-entry). Data given as the ratio of cross sections are transformed to $^{243}$Am(n, F) cross section using standard cross section obtained in this combined fit.

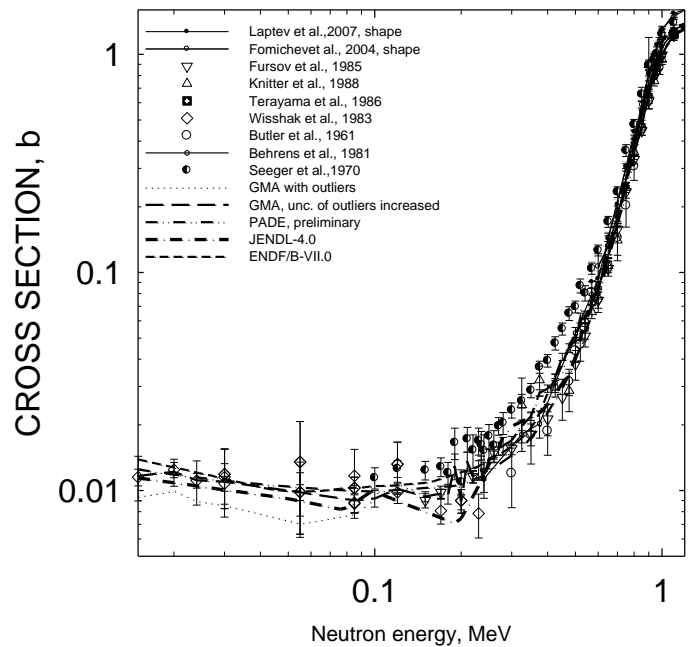

Fig. 3.4. The same as at Fig. 3.3 but for detailed view of the cross sections in the energy range from 0.015 to 0.8 MeV.

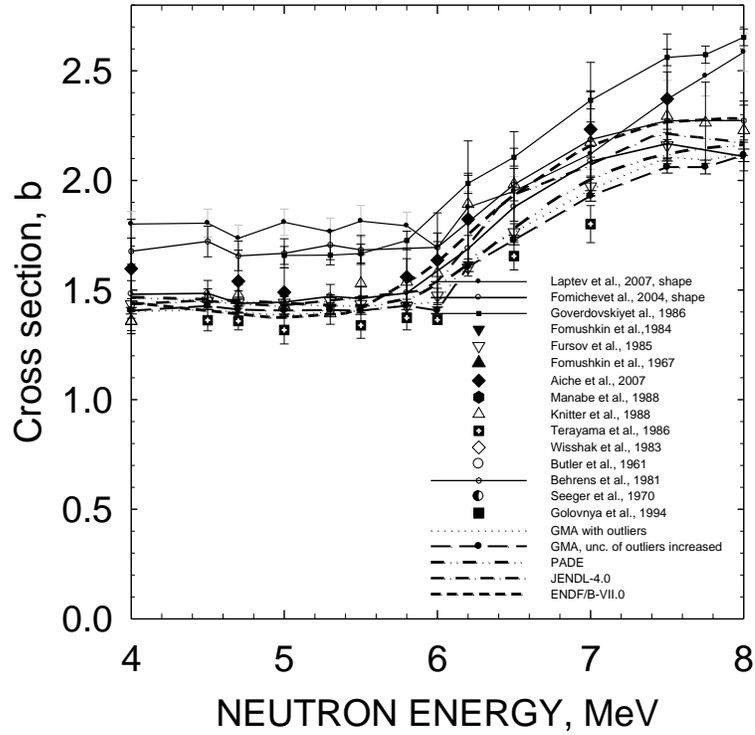

Fig. 3.4. The same as at Fig. 3 but for detailed view of the cross sections in the energy range from 4 to 7 MeV.

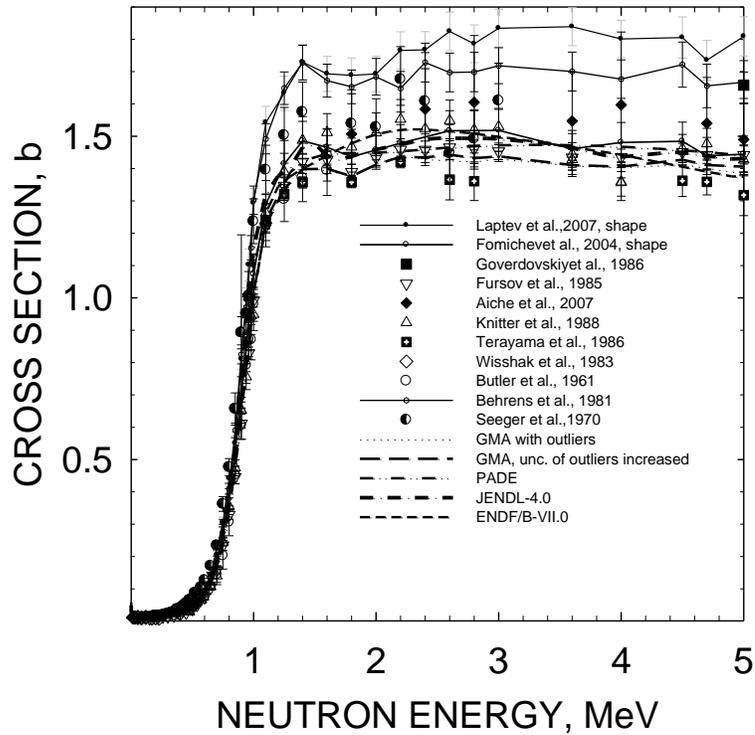

Fig. 3.6 Comparison of GMA' evaluated fission cross section of $^{243}$Am with previous evaluations and measured data.

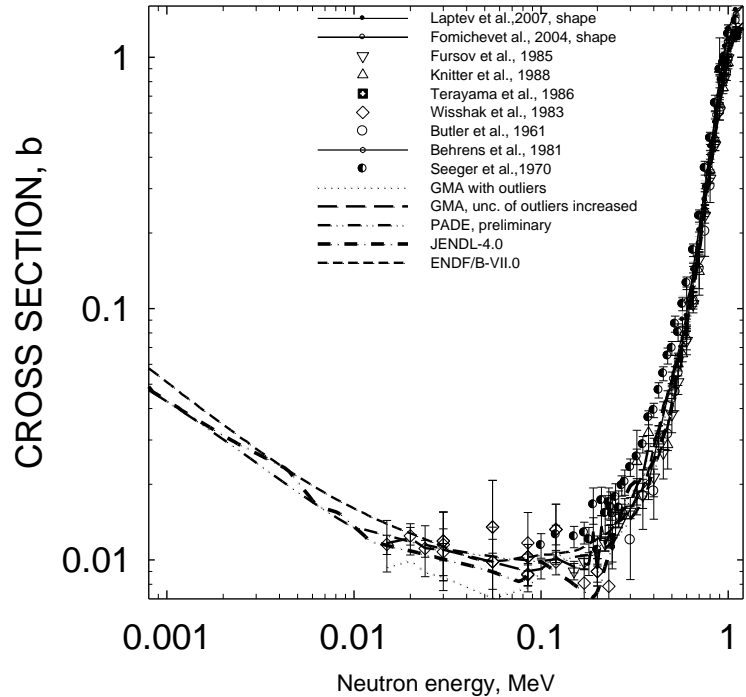

Fig. 3.7 Comparison of GMA' evaluated fission cross section of $^{243}$Am with previous evaluations and measured data.

The data presented at Fig. 3.3 — 3.7 show, that the experimental data have a rather large spread. GMA evaluation in the energy range above 1 MeV is below JENDL-4.0 evaluation, because of large influence of data by Terayama et al. [17] and lays slightly below data by Fursov et al. [22]. Largest differences with JENDL-4.0 evaluation were observed in the sub-threshold region, where uncertainties of the data are large.

As we see from figures 3.6 and 3.7, there is rather large spread of experimental data in the MeV region. New measurements are needed in the energy range 100 eV – 100 keV and in the 14 -20 MeV energy range.

## 4. Unresolved resonance parameters (0.250 - 96.8 keV)

Here we will briefly review the status of unresolved neutron resonance parameters of $^{243}$Am and provide a cross section parameterization of total, capture, elastic and inelastic scattering cross sections. The average resonance parameters were determined as described in [8] to reproduce average cross sections in the energy range of 0.25 keV-96.8 keV. Provided are energy dependent average resonance parameters.

We assume that the the upper energy is 96.8 keV, twice higher than in previous evaluations [8], the lower is 250 eV. We suppose s-, p- and d-wave neutron-nucleus interactions to be effective (see Tables 4.1).

**Table 4.1** Average neutron resonance parameters for $^{243}$Am

| | $D_{obs}$, eV | $\Gamma_\gamma$, meV | $S_o$ x $10^{-4}$ | $S_1$ x $10^{-4}$ | R, fm |
|---|---|---|---|---|---|
| JENDL-4.0 | .4356 | 39 | | | |
| | | | | | |
| RIPL | .730 | 39 | .98 | | |
| | | | | | |
| Present | 0.621 | 33 | .9009 | 2.205 | 9.236 |

### 4.1 Neutron resonance spacing

Neutron resonance spacing $D_{obs}$ was calculated with the phenomenological Ignatyuk' model [28], which takes into account the shell, pairing and collective effects. The main parameter of the model, asymptotic value of level density parameter $a$, was normalized to the observed neutron resonance spacing $D_{obs} = 0.621$ eV. Other parameters are provided in [29].

### 4.2 Neutron width

Average neutron width is calculated as follows

$$\left\langle \Gamma_n^{lJ} \right\rangle = S_l \left\langle D \right\rangle_J E_n^{1/2} P_l \nu_n^{lJ},$$ (4.1)

where $E_n$ is the incident neutron energy, $P_l$ is the transmission factor for the l-th partial wave, which was calculated within black nucleus model, $\nu_n^{lJ}$ is the number of degrees of freedom of the Porter-Thomas distribution. The p-wave neutron strength function $S_1 = 2.205 \times 10^{-4}$ at 250 eV was calculated with the optical model, using the deformed optical potential, described below. Figure 4.1 compares the average reduced neutron widths for the particular $(l, J)$- state, which is excited in the unresolved resonance energy range. The $\left\langle \Gamma_n^{0lJ} \right\rangle$ values for s-wave neutrons in ENDF/B-VII.0 [27] or JENDL-4.0 [13] data file are much less energy-dependent than those of present evaluation.

### 4.3 Radiative capture width

Energy and angular momentum dependence of radiative capture widths are calculated within a two-cascade γ-emission model with allowance for the (n, γn') [3] and (n, γf) [4, 31] reactions competition to the (n, γγ) reaction. The (n, γγ) reaction is supposed to be a radiation capture reaction. The radiation capture width was normalized to the value of $\Gamma_\gamma = 33$ meV, adopted here to describe the neutron capture cross section data. Detailed treatment is described below.

### 4.4 Neutron inelastic width

Average neutron inelastic width is calculated as follows

$$\left\langle \Gamma_{n'}^{lJ} \right\rangle_s = S_l \left\langle D \right\rangle_J (E_n - E')^{1/2} P_l (E_n - E') \nu_{n'}^{lJ},$$ (4.2)

where $\nu_{n'}^{lJ}$ is number of degrees of freedom of Porter-Thomas distribution. Excited levels of $^{243}$Am are taken from Nuclear Data Sheets [32].

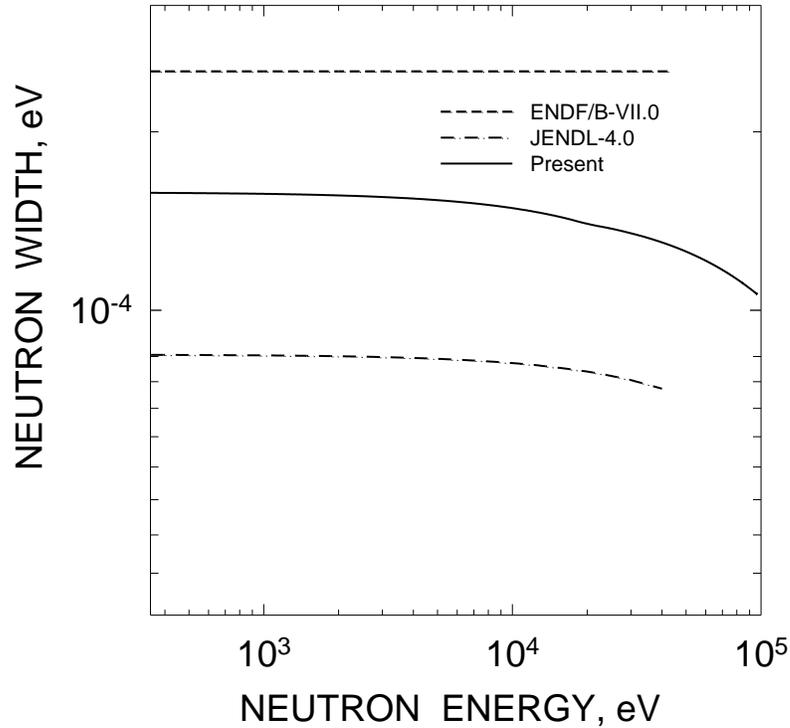

Fig. 4.1 Reduced neutron width of $^{243}$Am, $l$=0, $J$=2.

## 4.5 Fission width

Fission widths are calculated within a double-humped fission barrier model by Strutinsky [33]. Energy and angular momentum dependence of fission width is defined by the transition state spectra at inner and outer barrier humps as in [5]. We constructed transition spectra by supposing the triaxiality of inner saddle and mass asymmetry at outer saddle.

## 4.6 Total cross section in the region 0.25 keV-96.8 keV

Total cross section were calculated with the rigid rotator optical model. Coupled channel parameters, fitting the measured data are those defined for the $^{237}$Np+n interaction [5]. In the energy ranges 0.15 -20 keV and 20-96.8 keV the optical model calculations of the total cross section were reproduced assuming a decreasing trends of $S_o$, $S_1$ and $S_2$ strength function values as the latter and potential radius, which was adopted from the optical model calculations, define total cross section up to $E_n$= 93.998 keV.

To reproduce the $^{243}$Am total cross section, calculated with the optical model, we assume $S_o$ value linearly decreasing starting from 0.25 keV to $0.86885 \times 10^{-4}$, while $S_1$ decreases linearly to $1.9090 \times 10^{-4}$ at 96.8 keV (see Fig. 4.2). The d-wave neutron strength function was assumed to be equal to $S_2 = 1.0909 \times 10^{-4}$.

## 4.7 Elastic scattering cross section

The elastic scattering cross section is composed of shape elastic (see Fig. 4.3) and compound elastic contributions. Compound elastic scattering cross section estimate is rather insensitive to the $^{243}$Am fission cross section estimate. Present and JENDL-4.0 [13] estimates from 0.25 keV and up to 96.8 keV, shown on Fig. 4.3 differ a lot below 100 keV.

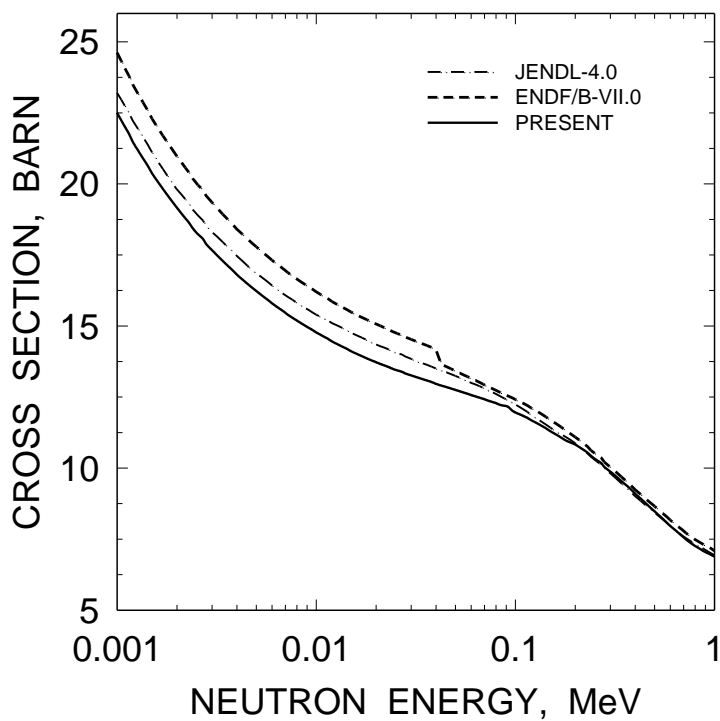

Fig. 4.2 Total cross section of $^{243}$Am.

## 4.8 Capture cross section

The description of capture cross section data by Weston et al. [11] and Wisshak et al. [12] in the energy range of 0.25 keV–200 keV is very sensitive to the shape of the neutron absorption cross section. It was shown in [5] that around $E_n \sim 1$ MeV the total cross section of Z-odd target nuclide is virtually insensitive to the lowering/increasing of the neutron absorption cross section. Lowering of the absorption cross section, simulated by the decrease of the imaginary surface potential term $W_D$, was cross-checked by the consistency of fission and inelastic scattering cross sections.

Measured data for the $^{243}$Am(n, $\gamma$) reaction cross section [11, 12] shown on Figs. 4.4, 4.5 are scattering a lot, or there are a systematic shifts between different data sets. At lower energy the GMA fits are inconsistent with statistical model calculations, defined by estimate of radiation strength function and neutron absorption cross section. Figure 4.5 shows the calculated curve, corresponding to the consistent description of the total, fission and inelastic scattering cross section with $<\Gamma\gamma>$ =33 meV and $<D_{obs}>$ =0.621 eV.

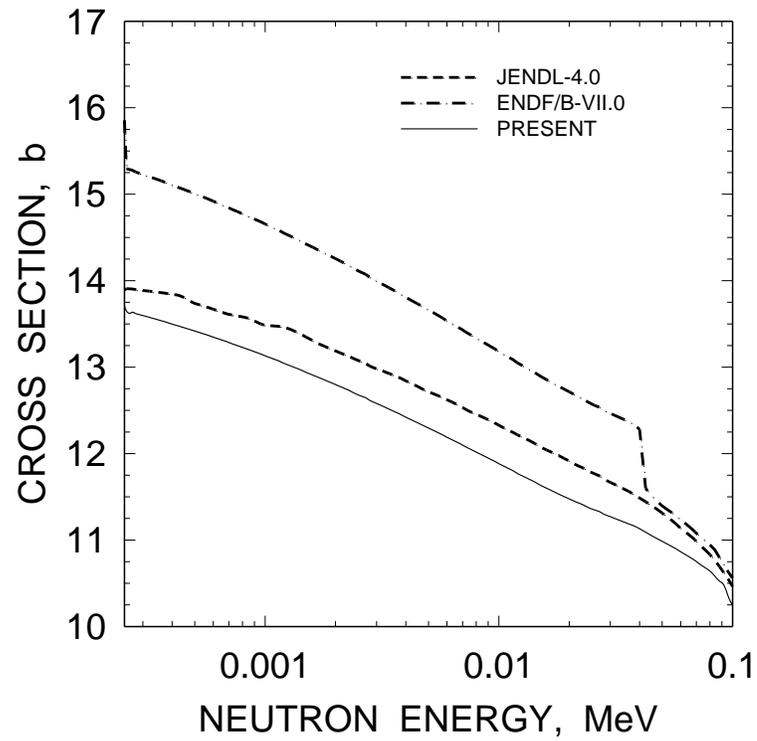

Fig. 4.3 Elastic cross section of $^{243}$Am.

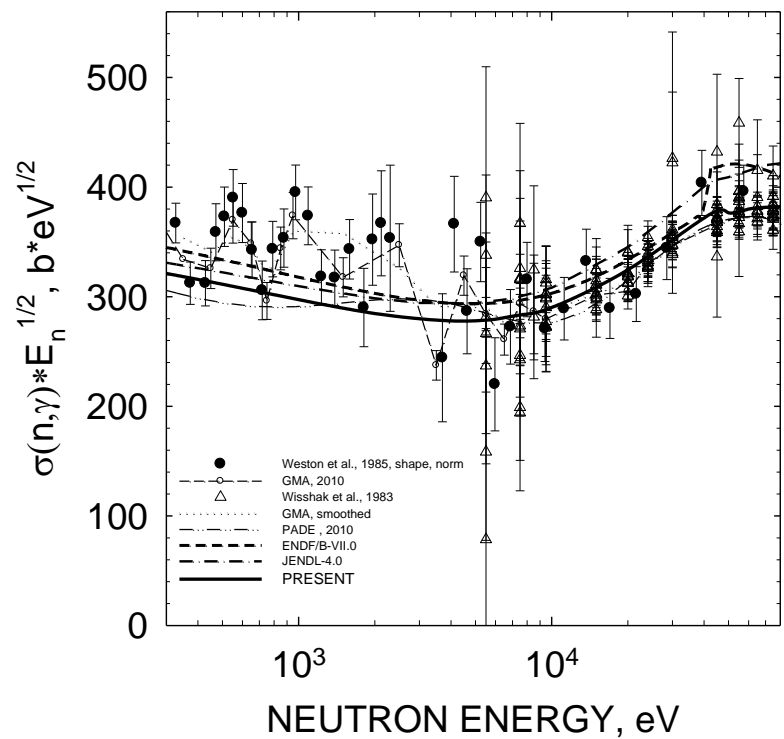

Fig. 4.4 Capture cross section of $^{243}$Am.

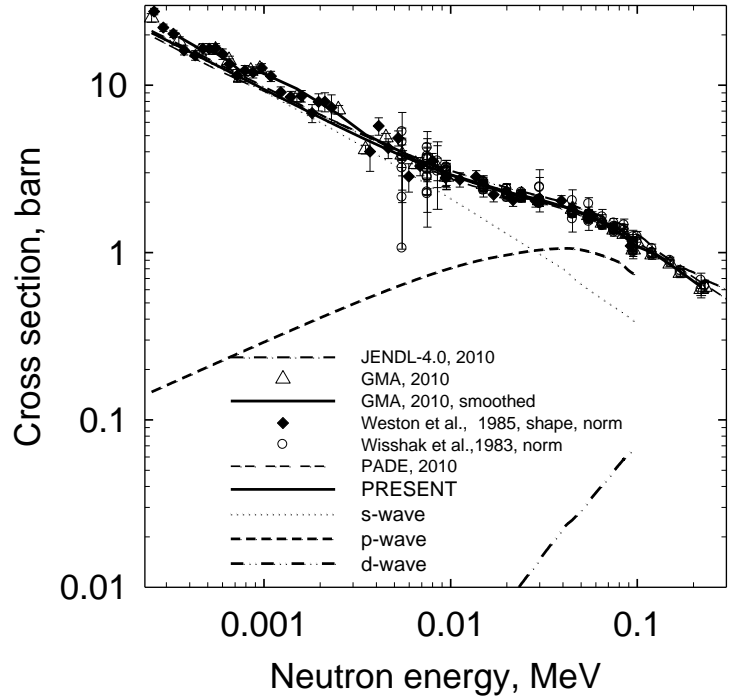

Fig. 4.5 Capture cross section of $^{243}$Am.

To follow a much higher lying data trend of the GMA-fit one would need rather large values of $s$-wave neutron strength function $S_0$.

The important peculiarity of the calculated $^{238}$U(n, $\gamma$) and $^{232}$Th(n, $\gamma$) [3, 4] capture cross sections, Wigner cusp above first excited level threshold, is pronounced in case of calculated $^{243}$Am(n, $\gamma$) reaction cross section differently, because of larger number of levels in odd residual $^{241}$Am nuclide. The pattern of $s$-, $p$- and $d$-wave entrance channel contributions to the capture cross section in the energy range of $0.25 - 96.8$ keV is different from that of $^{232}$Th [3] or $^{238}$U [4] target nuclides (see Fig. 4.5). That peculiarity may be traced to higher fissility of $^{244}$Am compound nuclide as well. In case of $^{238}$U(n, $\gamma$) reaction main contribution comes from p-wave neutrons above ~10 keV. The p-wave contribution to the $^{243}$Am(n, $\gamma$) reaction cross section is higher than that of $s$-wave above ~30 keV, while that of $d$-wave neutrons is the lowest.

### 4.9 Inelastic scattering cross section

Calculated inelastic scattering cross section is very close to previous evaluation of ENDF/B-VII.0 [30], but much different from that of JENDL-4.0 [13] (see Fig. 4.7). Conventional ENDF/B processing codes exemplify Hauser-Feshbach-Moldauer formalism. Figure 4.6 shows partial contributions to the inelastic scattering coming from different *(l, J)*-channels. Major contribution, unlike in case of $^{238}$U+n interaction [4], comes from $s$-wave channels (decay of 2$^-$ and 3$^-$ states), the intermediate comes from $p$-wave channel (decay of 0$^+$, 1$^+$, 2$^+$, 3$^+$, 4$^+$ compound nucleus states) and the lowest comes from $d$-wave neutrons (decay of 0$^-$, 1$^-$, 2$^-$, 3$^-$, 4$^-$, 5$^-$ compound nucleus states). At $E_n>$ 60 keV, the contribution from $p$-wave channel (decay of 0$^+$, 1$^+$, 2$^+$, 3$^+$, 4$^+$ compound nucleus states) becomes the largest, as shown on Fig. 4.7. However, that might be considered as imposed by the fitting procedures employed. Because Hauser-Feshbach-Moldauer formalism is adopted in conventional ENDF/B processing codes, the direct excitation of the 0.0424 MeV, $J^\pi=7/2^-$ level is not accounted for explicitly. To compensate for that relevant strength function $S_2$ for inelastic scattering exit channel was increased at 96.8 keV up to $2.989\times10^{-4}$. That helped to attain, using conventional processing codes, the fit of relevant capture cross section, calculated with the Hauser-Feshbach-Moldauer formalism.

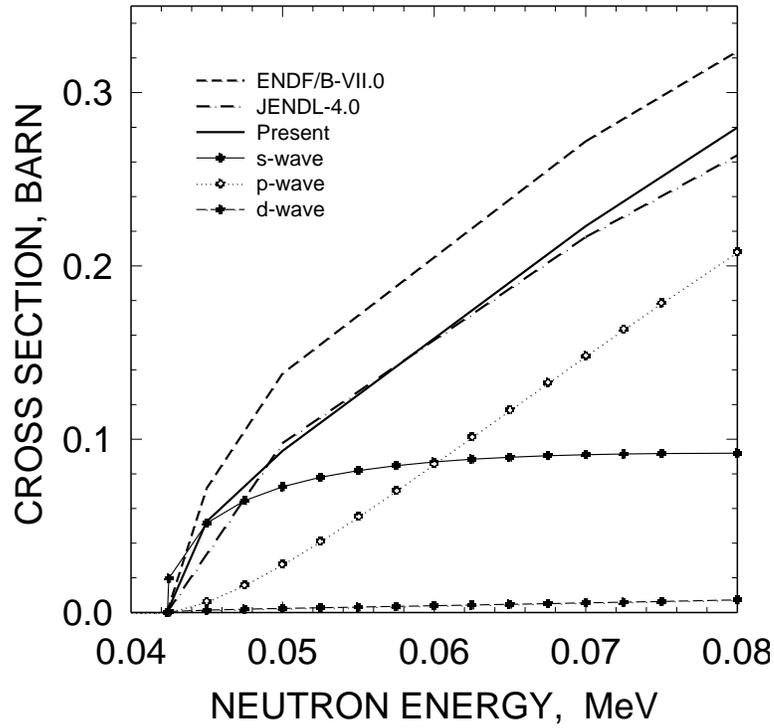

Fig. 4.6 Inelastic cross section of $^{243}$Am.

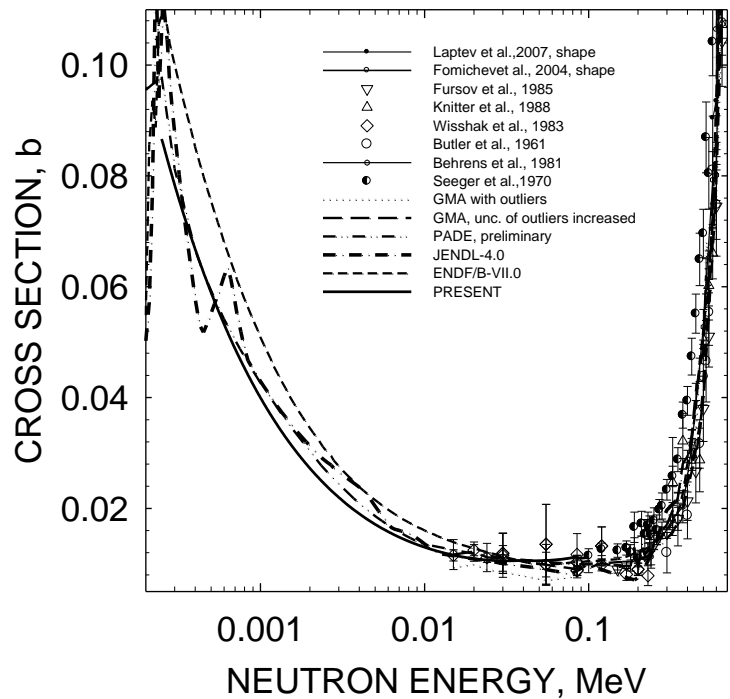

Fig. 4.7 Fission cross section of $^{243}$Am.

### 4.10 Fission cross section

Evaluated fission cross section describes the trend, predicted by data [14-27], specifically of Knitter et al. [18], Terayama et al. [17], Fursov et al. [22], Aiche et al. [20]. We estimated fission cross section in the unresolved resonance energy region using for transition state spectra of $^{244}$Am, fission barrier parameters were obtained fitting fission cross section data in the first plateau region (see Fig. 4.8). The fission cross section, calculated with the Hauser-Feshbach-Moldauer formalism, reproduces the GMA evaluation within assigned errors in the incident neutron energy range of 0.25 keV~96.8 keV.

### 5. Optical potential

Calculated total, elastic scattering and absorption cross sections were obtained with the coupled-channel potential parameters, obtained for the $^{237}$Np, as described in [5]. The experience of describing the capture cross section of $^{232}$Th [3] was the motivation to decrease the real volume potential term $V_R$ by 0.5 MeV. Rotational levels of ground state band $5/2^- - 7/2^- - 9/2^- - 11/2^-$ are assumed coupled (see Table 5.1).

Deformation parameters were tuned to fit $S_o$ and $S_1$ strength function values. Optical model potential parameters are defined as follows:

$V_R$=(45.7-0.334E) MeV;                     $r_R$ =1.2600 fm; $a_R$ =.6300 fm;
$W_D$=(3.690+0.400E)  MeV, $E_n$< 10 MeV          $r_D$ =1.24 fm;
$W_D$=  7.690  MeV, $E_n\geq$10 MeV              $a_D$ =.5200 fm;
$V_{SO}$= 6.2 MeV;                           $r_{SO}$=1.12 fm;
$a_{SO}$=.47 fm;                 $\beta_2$= 0.180              $\beta_4$=0.08.

**Table 5.1** $^{243}$Am level schema [32].

| No. | $E$ | $J^\pi$ |
|-----|--------|--------|
| 1 | 0.0 | 5/2$^-$ |
| 2 | 0.0422 | 7/2- |
| 3 | 0.084 | 5/2$^+$ |
| 4 | 0.0964 | 9/2- |
| 5 | 0.1092 | 7/2$^+$ |
| 6 | 0.1435 | 9/2$^+$ |
| 7 | 0.1623 | 11/2$^-$ |
| 8 | 0.1893 | 11/2$^+$ |
| 9 | 0.238 | 13/2$^-$ |
| 10 | 0.244 | 13/2$^+$ |
| 11 | 0.266 | 3/2$^-$ |

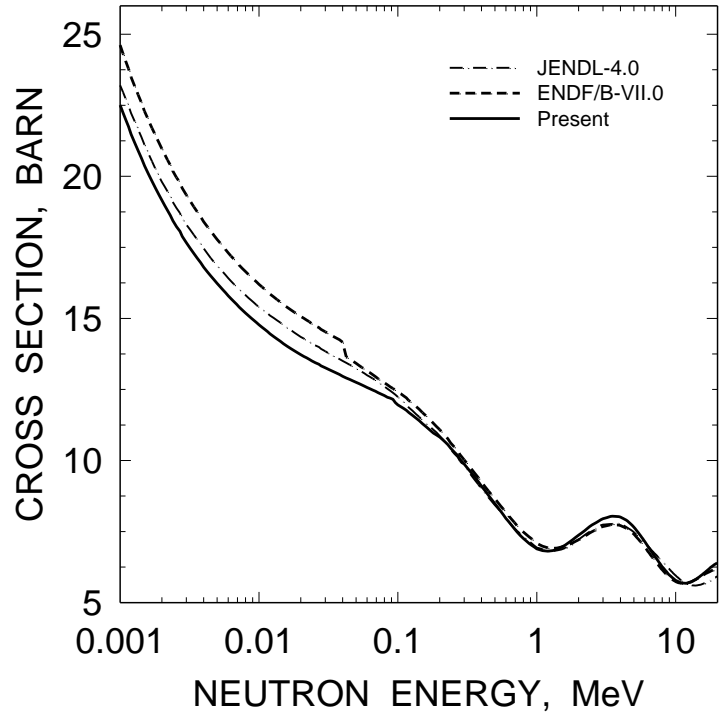

Fig. 6.1 Total cross section of $^{243}$Am.

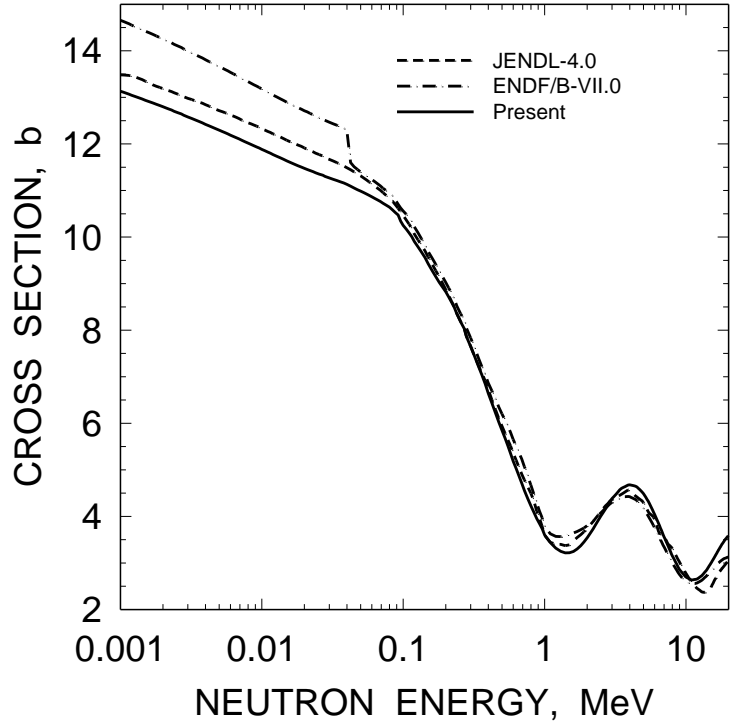

Fig. 6.2 Total and elastic cross section of $^{243}$Am.

Partitioning of the total cross section into absorption (reaction) and scattering cross sections allows get reasonable description of available fission and capture cross sections.

## 6. Total and elastic scattering cross section

Optical potential fitting Phillips and Howe [34] measured $^{241}$Am+n total cross section in the energy range of 0.4-25 MeV was adopted for $^{243}$Am+n. In case of ENDF/B-VII.0 [30] the elastic scattering was simply adjusted to balance total and partial cross sections (see Fig. 6.1).

## 7. Statistical model

We calculated neutron cross sections within Hauser-Feshbach theory, coupled channel optical model and double-humped fission barrier model.

Hauser-Feshbach-Moldauer [35] and Tepel-Hoffman-Weidenmuller [36] statistical theory is employed for partial cross section calculations below emissive fission threshold. Fissioning and residual nuclei level densities as well as fission barrier parameters are key ingredients, involved in actinide neutron-induced cross section calculations.

In case of fast neutron ($E_n \leq 6$ MeV) interaction with $^{243}$Am target nucleus, the main reaction channel is the fission reaction and fission cross section data description serves as a major constraint, except those of GMA-fits, for the neutron inelastic scattering and radiative neutron capture cross section estimates. Below there is an outline of the statistical model employed. At incident neutron energy lower than cut-off energy of discrete level spectra, neutron cross sections are calculated with Hauser-Feshbach formalism and width fluctuation correction by Moldauer [35]. For width fluctuation correction only Porter-Thomas fluctuations are taken into account. Effective number of degrees of freedom for fission channel is defined at the higher fission barrier saddle as $\nu_f^{J\pi} = T_f^{J\pi} / T_{fmax}^{JK\pi}$ where $T_{fmax}^{JK\pi}$ is the maximum value of the fission transmission coefficient. At higher incident neutron energies the Tepel et al. [36] approach is employed, it describes cross section behavior in case of large number of open channels correctly.

### 7.1 Level Density

Level density is the main ingredient of statistical model calculations. Level density of fissioning, residual and compound nuclei define transmission coefficients of fission, neutron scattering and radiative decay channels, respectively. The level densities were calculated with a phenomenological model by Ignatyuk et al. [28], which takes into account shell, pairing and collective effects in a consistent way

$$\rho(U, J^\pi) = K_{rot}(U, J) K_{vib}(U) \rho_{qp}(U, J^\pi), \qquad (7.1)$$

where quasi-particle level density is defined as

$$\rho_{qp}(U, J^\pi) = \frac{(2J+1)\omega_{qp}(U)}{4\sqrt{2\pi}\sigma_\perp^2 \sigma} \exp\left[-\frac{J(J+1)}{2\sigma_\perp^2}\right], \qquad (7.2)$$

$\rho_{qp}(U, J^\pi)$ is the state density, $K_{rot}(U, J)$ and $K_{vib}(U)$ are factors of rotational and vibrational enhancement of the level density. The closed-form expressions for thermodynamic temperature and other relevant equations which one needs to calculate $\rho(U, J^\pi)$ provided by Ignatyuk' model.

To calculate the residual nucleus level density at the low excitation energy, i.e. just above the last discrete level excitation energy where $N^{exp}(U) \approx N^{theor}(U)$, we employ a Gilbert-Cameron-type approach. The procedure is as follows. First, level density parameters are defined, using neutron resonance spacing $\langle D_{obs} \rangle$ estimate for $^{243}$Am target nuclide. Constant temperature level density parameters $T_o$, $E_o$, $U_c$ (see below for details) are defined by fitting cumulative number of low-lying levels of $^{243}$Am (see Fig. 7.1). Figure 7.2 shows the estimate of cumulative number of low-lying levels of $^{244}$Am, obtained using systematic of constant temperature level density parameters $T_o$, $E_o$, $U_c$ . On this figure levels of odd-odd $^{244}$Am nuclide are compared with constant temperature model estimate. The constant temperature approximation of the level density

$$\rho(U) = \frac{dN(U)}{dU} = \frac{1}{T} \exp\left( \frac{U - U_o}{T} \right) \tag{7.3}$$

is extrapolated up to the matching point $U_c$ to the $\rho(U)$ value, calculated with a phenomenological model by Ignatyuk with the condition

$$U_c = U_o - Tln(T\rho(\ U_c)). \tag{7.4}$$

In this approach $U_o \approx -m\Delta_o$, where $\Delta_o$ is the pairing correlation function, $\Delta_o = 12/\sqrt{A}$ , A is the mass number, m = 2 for odd-odd, 1 for odd-even nuclei, i.e. $U_o$ has the meaning of the odd-even energy shift. The value of nuclear temperature parameter $T$ obtained by the matching conditions of Eq. (7.4) at the excitation energy $U_c$.

The modeling of total level density

$$\rho(U) = K_{rot}(U,J) K_{vib}(U) \frac{\omega_{qp}(U)}{\sqrt{2\pi}\sigma} = \frac{1}{T} \exp\left( \frac{U - U_o}{T} \right) \tag{7.5}$$

in Gilbert-Cameron-type approach looks like a simple renormalization of quasi-particle state density $\omega_{qp}(U)$ at excitation energies U< $U_c$. The cumulative number of observed levels for odd-even $^{243}$Am and odd-odd nuclide $^{244}$Am [32] are compared with constant temperature approximations for $^{243}$Am and $^{244}$Am on Figs. 7.1 and 7.2, respectively. Missing of levels above ~0.5 MeV is pronounced in case $^{243}$Am.

Few-quasi-particle effects due to pairing correlations are essential for state density calculation at low intrinsic excitation energies only for equilibrium $^{243}$Am deformations. Few-quasi-particle effects in fissioning nuclide $^{244}$Am are unimportant because of its odd-odd nature.

The partial n-quasi-particle state densities for odd $^{243}$Am, which sum-up to intrinsic state density of quasi-particle excitations, could be modelled using the Bose-gas model prescriptions [37, 38]. The intrinsic state density of quasi-particle excitations $\omega_{qp}(U)$ could be represented as a sum of n-quasi-particle state densities $\omega_{nqp}(U)$:

$$\omega_{qp}(U) = \sum_n \omega_{nqp}(U) = \sum_n \frac{g^n (U - U_n)^{n-1}}{((n/2)!)^2 (n-1)!}, \tag{7.6}$$

where $g = 6a_{cr}/\pi^2$ is a single-particle state density at the Fermi surface, n is the number of quasi-particles. The important model parameters are threshold values $U_n$ for excitation of n-quasi-particle configurations employed, as applied for fission, inelastic scattering or capture reaction calculations, is provided in [39, 40].In case of and odd-odd nucleus $^{242}$Am Gilbert-Cameron-type approximation of $\rho(U)$ is employed. Nuclear level density $\rho(U)$ of odd nuclide $^{241}$Am at equilibrium deformation, as compared with the Gilbert-Cameron-type approximation of $\rho(U)$ is shown on Fig. 7.3. The arrows on the horizontal axis of Fig. 7.3 indicate the excitation thresholds of odd n-quasi-particle configurations.

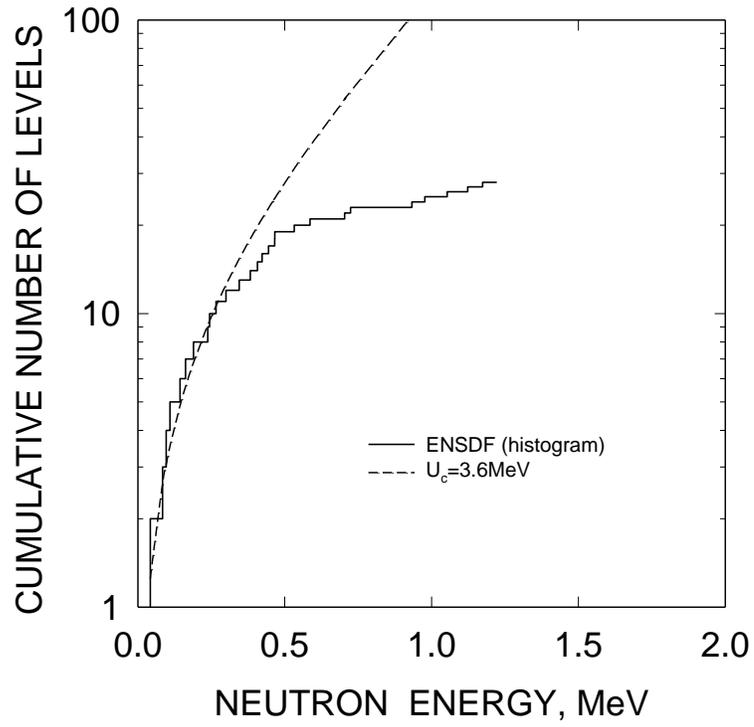

Fig. 7.1 Cumulative sum of levels of $^{243}$Am.

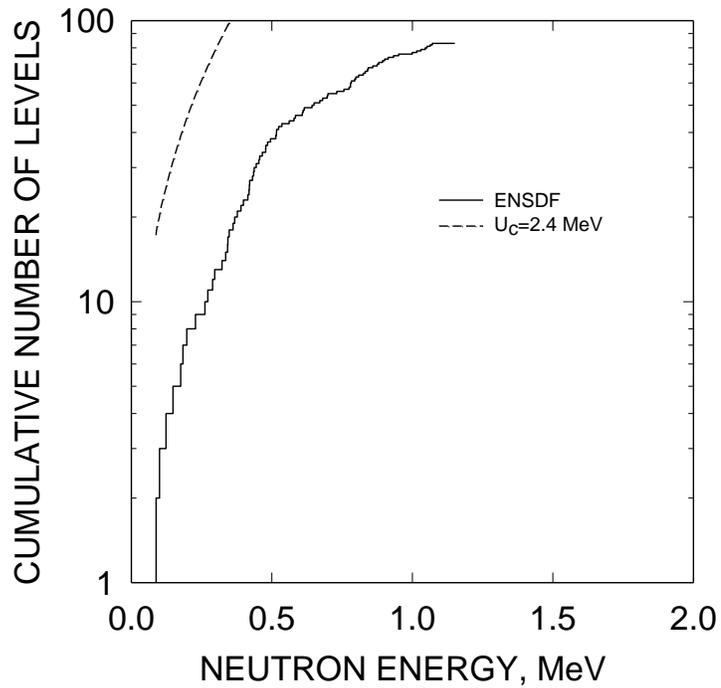

Fig. 7.2 Cumulative sum of levels of $^{244}$Am.

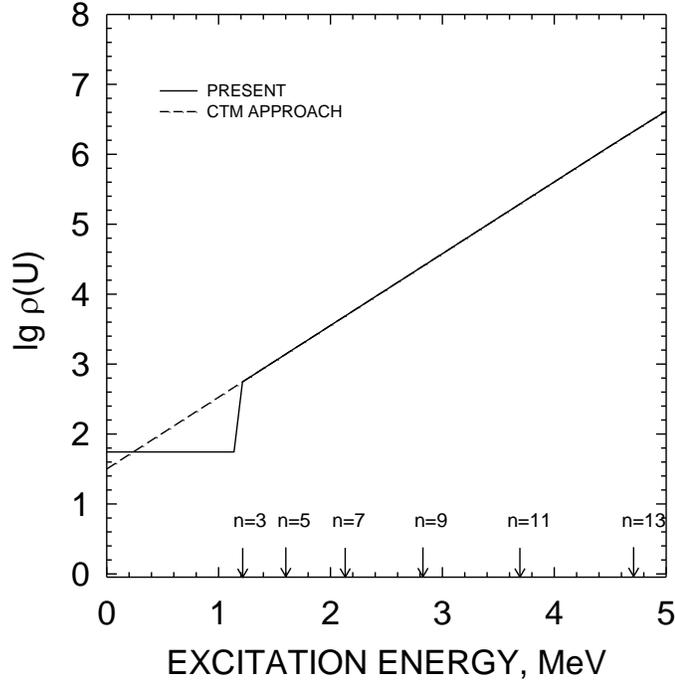

Fig. 7.3 Level density of $^{243}$Am.

Main parameters of the level density model for equilibrium, inner and outer saddle deformations are as follows: shell correction $\delta W$, pairing correlation functions $\Delta$ and $\Delta_f$, at equilibrium deformations $\Delta_o = 12/\sqrt{A}$, quadrupole deformation $\varepsilon$ and momentum of inertia at zero temperature $F_o/h^2$ are given in Table 7.1. For ground state deformations the shell corrections were calculated as $\delta W = M^{exp} - M^{MS}$, where $M^{MS}$ denotes liquid drop mass (LDM), calculated with Myers-Swiatecki parameters [41], and $M^{exp}$ is the experimental nuclear mass. Shell correction values at inner and outer saddle deformations $\delta W_f^{A(B)}$ are adopted following the comprehensive review by Bjornholm and Lynn [42].

**Table 7.1**. Level density parameters of fissioning nucleus and residual nucleus

| Parameter | Inner saddle | Outer saddle | Neutron channel |
|---|---|---|---|
| $\delta W$, MeV | 2.5* | 0.6 | LDM |
| $\Delta$, MeV | $\Delta_o + \delta$** | $\Delta_o + \delta$** | $\Delta_o$ |
| $\varepsilon$ | 0.6 | 0.8 | 0.24 |
| $F_o/h^2$, MeV | 100 | 200 | 73 |

*) for axially asymmetric deformations, 1.5 MeV for axially symmetric deformations;
**) $\delta = \Delta_f - \Delta_o$ value is defined by fitting fission cross section in the plateau region.

## 7.2 Fission cross section

Fission data, processed with GMA-code [9, 10], are used as a major constraint for capture, elastic and inelastic scattering, (n,2n) and (n,3n) cross sections as well as secondary neutron spectrum estimation. Description of measured fission cross section might justify a validity of level density description and fission barrier parameterization.

### 7.2.1 Fission Channel

Fission barrier of Am is double-humped, in the first "plateau" region and at higher energies we can use double-humped barrier model and relevant barrier parameters. Even at lower energies we could describe the general shape of the fission cross section starting from 0.250 keV.

Neutron-induced fission in a double humped fission barrier model is a two-step process, i.e. a successive crossing over the inner hump A and over the outer hump B. Hence, the transmission coefficient of the fission channel $T_f^{J\pi}(U)$ represented as

$$T_f^{J\pi}(U) = \frac{T_{fA}^{J\pi}(U)T_{fB}^{J\pi}(U)}{T_{fA}^{J\pi}(U) + T_{fB}^{J\pi}(U)}. \tag{7.7}$$

The transmission coefficient $T_{fi}^{J\pi}(U)$ is defined by the level density $\rho_{fi}(\varepsilon, J, \pi)$ of the fissioning nucleus at the inner and outer humps (i = A, B, respectively):

$$T_{fi}^{J\pi}(U) = \sum_{K=-J}^{J} T_{fi}^{JK\pi}(U) + \int_0^U \frac{\rho_{fi}(\varepsilon, J, \pi)d\varepsilon}{1 + \exp\left(2\pi\left(E_{fi} + \varepsilon - U\right)/\hbar\omega_i\right)}, \tag{7.8}$$

where the first term denotes the contribution of low-lying collective states and the second term that coming from the continuum levels at the saddle deformations, $\varepsilon$ is the intrinsic excitation energy of fissioning nucleus. The first term contribution due to discrete transition states depends upon saddle symmetry. The total level density $\rho_{fi}(\varepsilon, J, \pi)$ of the fissioning nucleus is determined by the order of symmetry of nuclear saddle deformation.

Inner and outer fission barrier heights and curvatures as well as level densities at both saddles are the model parameters. They are defined by fitting fission cross section data at incident neutron energies below emissive fission threshold. Fission barrier height values and saddle order of symmetry are interdependent. The symmetry of nuclear shape at saddles was defined by Howard and Moller [43] within shell correction method (SCM) calculation. We adopt the saddle point asymmetries from SCM calculations. According to shell correction method (SCM) calculations of Howard and Moller [43] the inner barriers were assumed axially asymmetric. Outer barriers for the americium nuclei are assumed mass-asymmetric.

### 7.2.2 Fission transmission coefficient, level density and transition state spectrum

Adopted level density description allows describe shape of measured fission cross section data of $^{243}$Am (see Figs. 7.4-7.7). One- and two-quasi-particle states in odd residual nuclide $^{243}$Am could be excited. The transition state spectra of odd-odd $^{244}$Am nuclide for the band-heads of Table 7.2 were constructed using values of $F_o/\hbar^2$ at the inner and outer saddles shown in Table 7.1.

We construct the discrete transition spectra up to 175 keV, using collective states of Table 7.2. The discrete transition spectra, as well as continuous level contribution to the fission

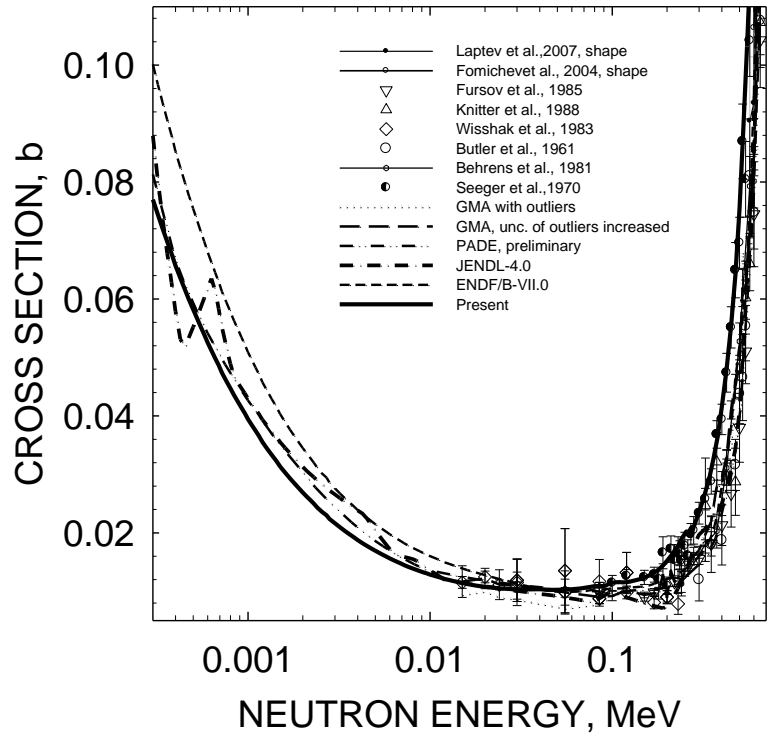

Fig.7.4 Fission cross section of $^{243}$Am.

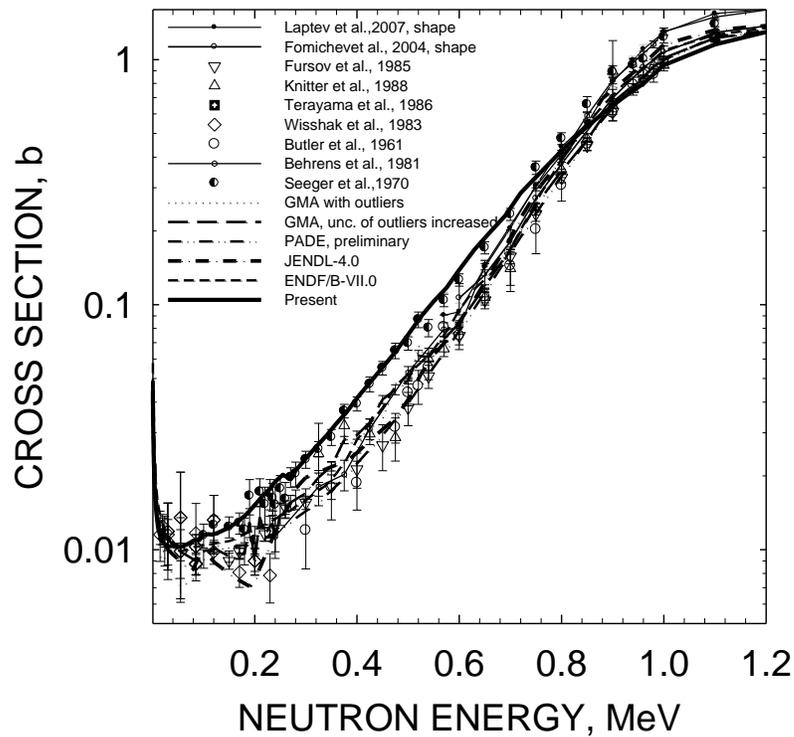

Fig. 7.5 Fission cross section of $^{243}$Am.

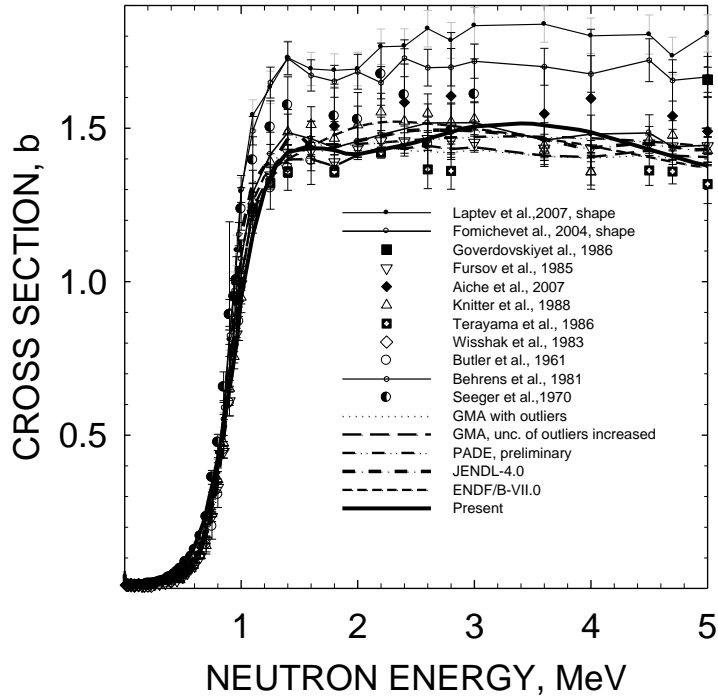

Fig. 7.6 Fission cross section of $^{243}$Am.

We construct the discrete transition spectra up to 175 keV, using collective states of Table 7.2. The discrete transition spectra, as well as continuous level contribution to the fission transmission coefficient are dependent upon the order of symmetry for fissioning nucleus at inner and outer saddles. With transition state spectra defined as described, the fission barrier parameters are obtained.

**Table 7.2** Transition spectra band-heads, Z-odd, N-even nuclei

| Inner saddle A | | Outer saddle B | |
|---|---|---|---|
| $K^\pi$ | $E_{K\pi}$, MeV | $K^\pi$ | $E_{K\pi}$, MeV |
| $2^+$ | 0 | $2^+$ | 0 |
| $3^+$ | 0.05 | $3^+$ | 0.05 |
| $3^-$ | 0.05 | $3^-$ | 0.05 |
| $2^-$ | 0.05 | $2^-$ | 0.05 |

## 7.3 Fission data analysis

Fission cross section is calculated within statistical model from 0.25 keV up to the emissive fission threshold. Measured fission data [14-27] analysis was accomplished within GMA approach [9, 10], as described above. Calculated cross section is consistent with data by Wsshak et al. [12], Knitter et al. [18], Terayama et al. [17], Fursov et al. [22], Fomushkin et al. [21, 23], Aiche et al. [20], Seeger et al. [44].

Statistical model calculations in the energy of .25 keV~6 MeV are maintained, calculated cross sections deviate from the GMA evaluation within the GMA-estimated uncertainties, except part of the threshold range.

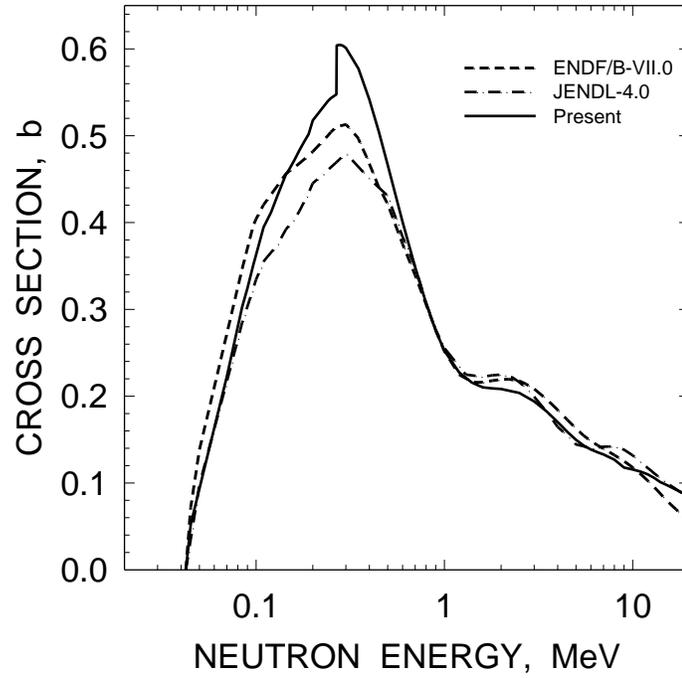

Fig. 7.8 Inelastic cross section of 1$^{st}$ level of $^{243}$Am.

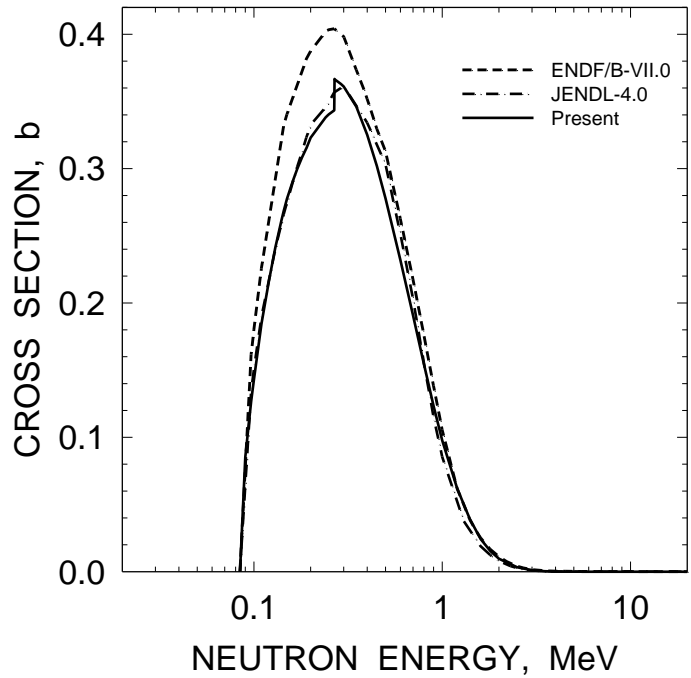

Fig. 7.9 Inelastic cross section of 2$^{nd}$ level of $^{243}$Am.

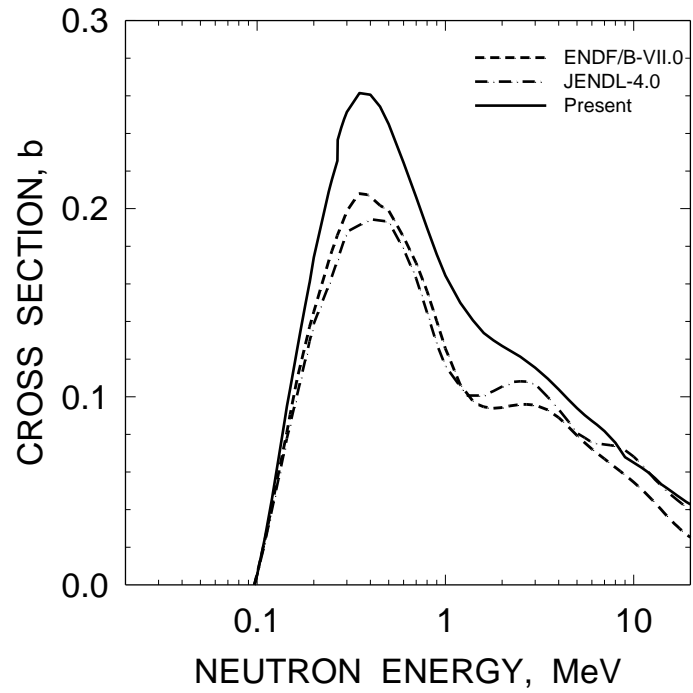

Fig. 7.10 Inelastic cross section of 3$^{rd}$ level of $^{243}$Am.

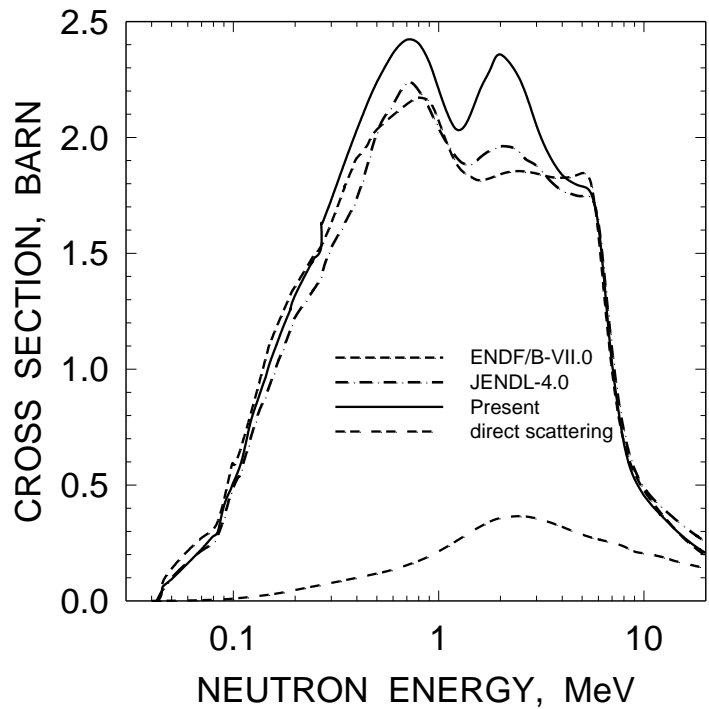

Fig. 7.11 Inelastic cross section of $^{243}$Am.

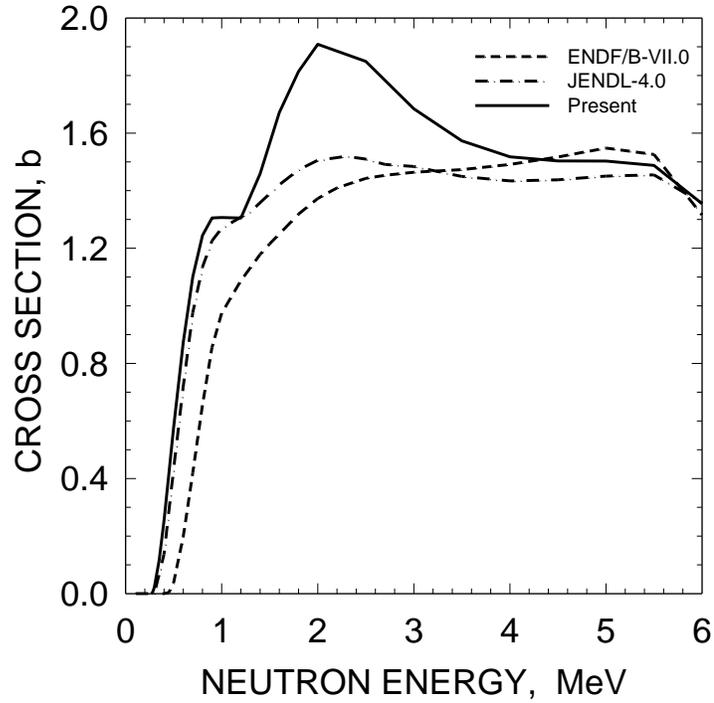

Fig. 7.12 Inelastic cross section of $^{243}$Am(continuum contribution)..

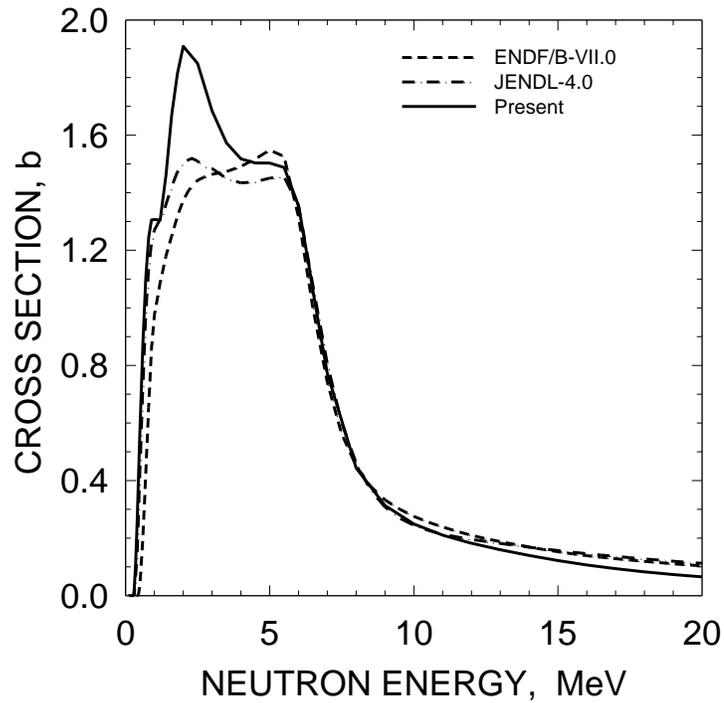

Fig. 7.13 Inelastic cross section of $^{243}$Am(continuum contribution)..

We fit the trend of the fission cross section data above $E_n \approx 3$ MeV increasing the correlation function value at outer saddle, which controls the $^{243}$Am (n, f) cross section shape. For incident neutron energies up to $E_n \approx 3$ MeV the threshold shape is roughly reproduced by varying the density of one-quasi-particle states of the residual nuclide $^{243}$Am (see Figs. 7.4-7.6). Smooth statistical model calculations are adopted as evaluated fission cross section in the energy range of 0.15 keV~5.5 MeV.

## 7.4 Inelastic scattering

Fission data fit largely defines the compound inelastic neutron scattering contribution to the total inelastic scattering cross section. The relative contribution of direct discrete level excitation cross sections is much higher than in case of say $^{238}$U target nuclide due to much higher fission competition to the compound neutron scattering. That explains the sensitivity of the $^{243}$Am compound inelastic scattering cross section to the fission competition and modeling of the residual nuclide level density.

### 7.4.1 Neutron Channel

The lumped transmission coefficient of the neutron scattering channel is given by equation

$$T_n^{J\pi}(U) = \sum_{l'j'q'} T_{l'j'}^{J\pi}(E - E_{q'}) + \sum_{l'j'I'} \int_0^{U-U_c} T_{l'j'}^{J\pi}(E')\rho(U - E', I'^\pi)dE', \qquad (7.9)$$

where $\rho(U - E', I'^\pi)$ is the level density of the residual nucleus. Levels of residual nuclide $^{241}$Am are provided in Table 5.1. The entrance channel neutron transmission coefficients $T_{l'j'}^{J\pi}(E')$ are calculated within a rigid rotator coupled channel approach. The compound and direct inelastic scattering components summed incoherently. The exit channel neutron transmission coefficients $T_{l'j'}^{J\pi}(E')$ were calculated using the re-normalized deformed optical potential of entrance channel without coupling, which describes a neutron absorption cross section.

### 7.4.2 Ground State Rotational Band

Predicted discrete level excitation cross section shape, calculated within a rigid rotator model, depends upon optical potential used. We assume strong missing of levels above excitations of 0.313 MeV (see Fig. 7.1), so only 10 excited levels up to this excitation energy were included when calculating inelastic scattering cross sections. Predicted discrete level excitation cross section shape, calculated within a rigid rotator model, strongly depends upon optical potential used. Calculated compound contribution is controlled mainly by fission competition (see Figs. 7.8-7.13). Figures 7.8, and 7.10 show that direct scattering essentially defines the excitation cross section of $J^\pi=7/2^-$ and $J^\pi=9/2^-$ levels of the ground state band levels at $E_n \geq 1$ MeV. Discrepancies with previous evaluations are due to both compound and direct contributions differences. The compound component tends to be zero above incident neutron energy of ~3 MeV (see Fig. 7.9).

### 7.4.3 Total inelastic cross section

Direct inelastic contributions added incoherently to Hauser-Feshbach calculated values of compound nucleus inelastic scattering cross sections. Present calculation based on the fits of the fission cross section. The evaluated inelastic cross sections of ENDF/B-VII.0 [30] and JENDL-4.0 [13] evaluations are in severe disagreement with our evaluation in the energy range 0.30 MeV − 2 MeV) (see Fig. 7.11).

Upward trend of the inelastic data at $E_n \geq 1.5$ MeV might be explained by the sharp increase of the level density of the residual nuclide $^{243}$Am due to the onset of three-quasi-particle excitations [37, 38], that conclusion is qualitatively supported by the measured data by Kornilov et al [45] for $^{237}$Np target nuclide [5] (see Figs. 7.11-7.13). The total inelastic scattering cross section of JENDL-4.0 [13] is much lower, than present evaluation. The continuum levels contribution to the total inelastic scattering cross section is shown on Figs. 7.12, 7.13.

## 7.5 Capture cross sections

We have demonstrated by the analysis of measured capture cross sections of $^{238}$U(n, γ), $^{232}$Th(n, γ) and $^{237}$Np(n, γ) [3, 4, 5] that neutron capture data could be described within a Hauser-Feshbach-Moldauer [35] statistical model, reproducing delicate variations of the measured cross sections with the increase of the incident neutron energies. Specifically, in a few-keV energy region calculated capture cross section is defined by the radiative strength function value $S_\gamma = \Gamma_\gamma/D$. At incident neutron energies above $E_n \approx 100$ keV calculated capture cross section shape is defined by the energy dependence of the radiative strength function $S_\gamma$. Energy dependence of $S_\gamma$ is controlled mainly by the energy dependence of the level density of the compound nuclide $^{244}$Am. Alongside with neutron emission at the second γ-cascade rather low fission threshold for the $^{244}$Am nuclide defines appreciable competition of fission [31], i.e. after first γ-quanta emission.

Then "true" capture reaction cross section (n,γγ) is defined using transmission coefficient $T_{\gamma\gamma}^{J\pi}(E)$, which is defined in a two-cascade approximation as

$$T_{\gamma\gamma}^{J\pi}(E) = \frac{2\pi C_{\gamma1}}{3(\pi\hbar c)^2} \int_0^{E_n+B_n} \varepsilon_\gamma^2 \sigma_\gamma(\varepsilon_\gamma) \sum_{I=|J-1|}^{I=J+1} \rho(U-\varepsilon_\gamma, I, \pi) \frac{T_\gamma^{1\pi}(U)}{T_f^{1\pi}(U) + T_{n'}^{1\pi}(U) + T_\gamma^{1\pi}(U)} d\varepsilon_\gamma . \quad (7.10)$$

The last term of the integrand describes the competition of fission, neutron emission and γ-emission at excitation energy $(U - \varepsilon_\gamma)$ after emission of first γ-quanta, $C_{\gamma1}$ is the normalizing coefficient. That means that transmission coefficients $T_f^{1\pi}(U), T_{n'}^{1\pi}(U), T_\gamma^{1\pi}(U)$ are defined at excitation energy $(U - \varepsilon_\gamma)$. The neutron emission after emission of first γ-quanta strongly depends on the $^{241}$Am residual nuclide level density at excitations around the pair-breaking threshold in odd nuclide $U_3$. The contribution of (n, γf)-reaction to the fission cross section is defined by $T_{\gamma f}^{J\pi}(E)$ value. The energy dependence of (n γf) reaction transmission coefficient $T_{\gamma f}^{J\pi}(E)$ was calculated with the expression

$$T_{\gamma f}^{J\pi}(E) = \frac{2\pi C_{\gamma1}}{3(\pi\hbar c)^2} \int_0^{E_n+B_n} \varepsilon_\gamma^2 \sigma_\gamma(\varepsilon_\gamma) \sum_{I=|J-1|}^{I=J+1} \rho(U-\varepsilon_y, I, \pi) \frac{T_f^{1\pi}(U)}{T_f^{1\pi}(U) + T_{n'}^{1\pi}(U) + T_\gamma^{1\pi}(U)} d\varepsilon_\gamma . \quad (7.11)$$

Competition of (n, γn') reaction is taken into account in a similar way. Above neutron energy 5.5 MeV capture cross section is assumed to be 0.001 barn.

Trends of the measured data by Weston et al. [11] and Wisshak et al. [12] are inconsistent with each other. Measured data for the $^{243}$Am(n, γ) reaction cross section [11, 12] shown on Figs. 3.1, 3.2 are scattering a lot, or there are a systematic shifts between different data sets. The GMA-fit follows the data by Weston et al. [11]. In the incident neutron energy range of 20-300 keV, the calculated capture cross section closely follows the GMA-fit. The trend predicted by the data of Wisshak et al. [12] is divergent with the former one at $E_n < 20$ keV. However, the data in the range of 10~100 keV support the theoretical calculation, based on consistent description of fission measured data, as well as the estimate of radiation strength function ($<\Gamma_\gamma> = 33$ meV and $<D_{obs}> = 0.621$ eV) and absorption cross section (see Figs. 7.14, 7.15).

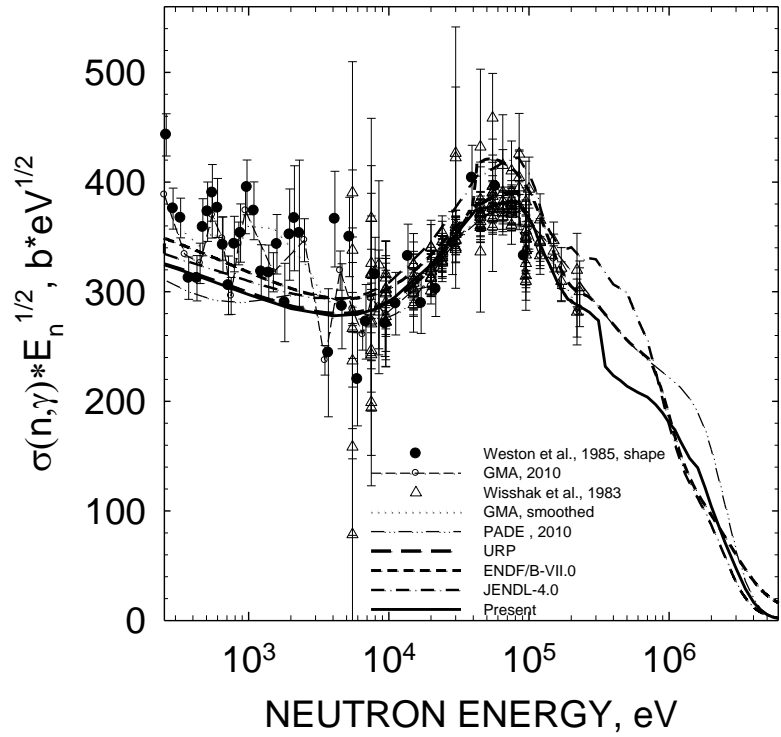

Fig. 7.14 Capture cross section of $^{243}$Am

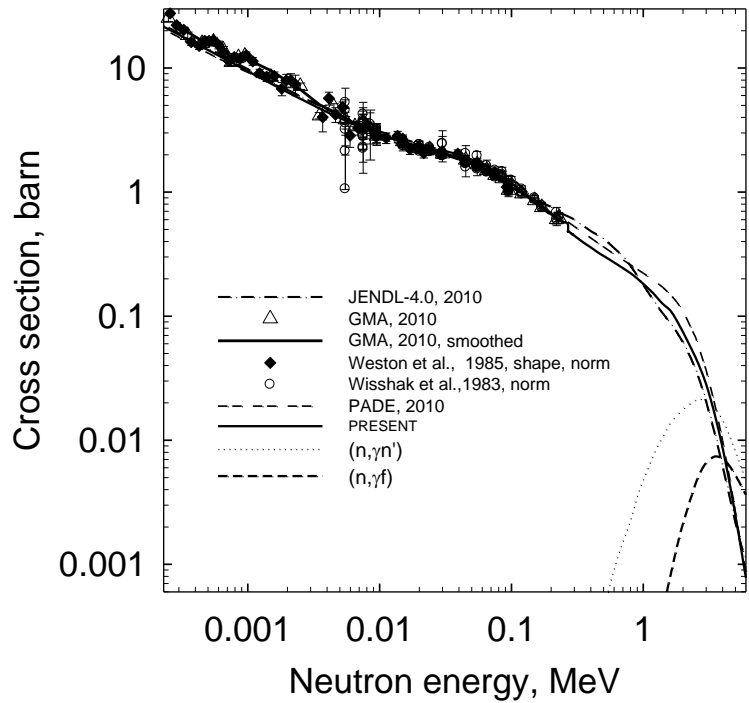

Fig. 7.15 Capture cross section of $^{243}$Am

Finally, Fig. 7.15 shows calculated capture cross section and competition of $^{243}$Am(n, γf) and $^{243}$Am (n, γn') reactions to the "true" capture reaction $^{243}$Am(n, γ γ), they define the capture cross section shape at $E_n \geq 2$ MeV.

## 7.6 Branching ratio of short-lived $^{244m}$Am (1⁻) and long-lived $^{244g}$Am (6⁻) states of $^{244}$Am

The neutron capture reaction $^{243}$Am(n, γ) populates either the 10.1-h ground state $^{244g}$Am or the $^{244m}$Am isomer with $T_{1/2} = 26$ min. Experimental information on (n,γ) reaction excitation function for the target actinide $^{243}$Am [1, 11, 12, 46-60] is devoted to measurements of thermal capture cross sections. Only in two experiments [11, 12] the $^{243}$Am(n, γ)$^{244m+g}$Am reaction excitation function was investigated in the energy range of 0.258 – 226 keV. We analyzed and corrected experimental cross section data to modern standards and evaluated the isomeric cross section ratio for the yields of metastable state ($t_{1/2}$=26 min) $^{244m}$Am, having spin $J^\pi$=1⁻ and ground state ($t_{1/2}$=10.1 h) state $^{244g}$Am, having spin $J^\pi$=6⁻: $^{243}$Am(n, γ)$^{244g}$Am/$^{243}$Am(n, γ)$^{244m+g}$Am.

All reviewed experimental data, if possible, were renormalized to the new standards for the decay data and cross sections for the monitor reactions. Direct measurements of the $^{243}$Am(n,γ)$^{244m}$Am reaction cross-section is not feasible. Analysis of the experiments for determination of the $^{243}$Am thermal capture cross-sections show, that most representative are experimental data by Gavrilov et al. [56] and Marie et al. [1]. Evaluated $^{243}$Am thermal capture cross-sections are given below in Table 7.1.

Table 7.1. Evaluated the $^{243}$Am(n, γ)$^{244m}$Am, $^{243}$Am(n, γ)$^{244g}$Am and $^{243}$Am(n, γ)$^{244m(g)}$Am cross sections at $E_n = 0.0253$ eV

| Cross-section | Evaluated value, b |
|---|---|
| $\sigma^g$ | 5.2 ± 31.62% |
| $\sigma^m$ | 76.39 ± 6.15% |
| $\sigma^{m+g}$ | 81.59 ± 3.75% |

The isomeric cross-section ratio $\sigma^g(E_n)/\sigma^{m+g}(E_n)$ for the reaction $^{243}$Am(n,γ) may be determined from the evaluated experimental data at the thermal point 0.0253 eV and in the resonance energy range between 0.5 eV and about 1 MeV on the basis of experimental data by Schuman [52].

Calculated from the evaluated experimental data (Table 7.1 ) isomeric cross-section ratio at 0.0253 eV is equal to $\sigma^g$(0.0253 eV)/$\sigma^{m+g}$(0.0253 eV)= 0.064 ± 7.20%. The isomeric cross-section ratio obtained in the resonance energy range from experimental data by Schuman [52] equals to $\sigma^g(E_n)/\sigma^{m+g}(E_n)$= 0.051 ± 13.07%, there $E_n$ ranges from 0.5 eV to ~1 MeV. It is necessary to mention, that measured by Schuman [52] thermal caprure cross section $\sigma^g$(0.0253 eV) = 5.9 b ± 33% agrees satisfactory with new experimental data by Marie et al. [1] $\sigma^g$(0.0253 eV) = 5.2 b ± 31.62%.

The neutron capture reaction $^{243}$Am(n, γ) populates either the $T_{1/2}$ =10.1h ground state $^{244g}$Am with $J^\pi$=1⁻ or the $^{242m}$Am isomer $J^\pi$=5⁻ with $T_{1/2}$ =26 min. The ground state $^{244g}$Am β⁻ −decays to $^{244}$Cm via $^{242m}$Am(n, γ)$^{243}$Am(n,γ)$^{244g}$Am(β⁻)$^{244}$Cm. It gives a path for the $^{244}$Cm yield also via $^{242m}$Am(n, γ)$^{243}$Am(n, γ)$^{244m}$Am(β⁻(ε))$^{244}$Cm($^{244}$Pu).

Ground $^{242g}$Am and isomer $^{242m}$Am states of the residual nuclide $^{242}$Am are excited in the reaction $^{243}$Am(n, 2n)$^{242g(m)}$Am as well. The same approach as in case of capture reaction is applied to predict the branching ratio $r(E_n) = \sigma^g_{n2n}(E_n)/\sigma^m_{n2n}(E_n)$ in $^{237}$Np(n, 2n) reaction, as described in [5]. The branching ratio $r(E_n) = \sigma^g_\gamma(E_n)/\sigma^m_\gamma(E_n)$ from thermal energy to 20 MeV could be defined by the ratio of the populations of two lowest states in $^{244}$Am [61]. These populations are defined by the γ-decay of the excited states, described by the kinetic equation, developed by Strutinsky et al. [62]. The branching ratio $r(E_n)$ is defined by the ratio of the populations of the two lowest states, $^{244g}$Am, with spin $J = 6$ and $^{244m}$Am, with spin $J = 1$. The integral equation in the code STAPRE [63] solved as a system of linear equations, the integration range is binned, in the assumption that there are no γ-transitions inside the bins as described in [5].

The isomer branching ratio depends mostly on the low-lying levels scheme and relevant γ-transitions probabilities. Though experimental data are available for $^{244}$Am [32], we will use a simplified approach, since level scheme and gamma-decay branching ratios are still incomplete. Modeling of low-lying levels of $^{244}$Am in [61] is accomplished based on the assumption that ground and first few excited states are of two-quasi-particle nature. For actinides with quadrupole deformations the superposition principle is usually adopted, the band-head energies of the doubly-odd nucleus are generated by adding to the each unpaired configuration $(\Omega_p, \Omega_n)$, as observed in the isotopic/isotonic (A-1) nucleus, the rotational energy contribution and residual (n−p) interaction energy contribution. The angular momenta of neutron and proton quasi-particles could be parallel or anti-parallel. In the independent quasi-particle model the two-quasi-particle states, $K^+ = |K_n + K_p|$ and $K^- = |K_n - K_p|$, are degenerate. Gallaher-Moshkowski doublets [61] appear because of the (n−p) residual interaction. Figure 7.16 shows employed band-head energies for the two-quasi-particle states expected in the odd-odd nuclide $^{244}$Am up to 700 keV. The spectroscopic properties of two pairs of proton and neutron single particle states were derived from those experimentally observed in the isotopic (Z=95) and isotonic (N=147) odd-mass nuclei with mass (A-1). Figure 7.16 shows levels expected, which have similar ordering as experimentally observed [32]. For the band-heads, shown on Figure 7.16, the rotational bands were generated as

$$E_{JK\pi} = E_{JK} + 5.5[J(J+1) - K(K+1)]. \tag{7.10}$$

Obviously, the schema presented on Fig. 7.16 does not represent a complete set to allow the calculation of absolute yields of $^{243}$Am (n,γ) $^{244m}$Am and $^{243}$Am (n,γ) $^{244g}$Am reactions. Rotational bands were constructed up to 700 keV excitation, modeling levels with spins $J^\pi \leq 10$, in total up to 70 levels. It was shown in [23], that simple estimate of the number of levels in odd-odd nuclei as

$$N(U) = e^{2\Delta_0/T}(e^{U/T} - 1), \tag{7.11}$$

predicts up to 280 level at $U \sim 700$ keV, $T = 0.388$ MeV, $\Delta = 12/A^{1/2}$, MeV. We assume that the modeled angular momentum distribution would not be much different from realistic estimates. Since the complete data on the γ-transitions are missing, we assumed the simple decay scheme: only E1, E2 and M1, M2, M3 transitions are allowed in a continuum excitation energy range. Inter-band transitions are not allowed, i.e., only γ-transitions within the rotational bands are possible. In that approach the populations of the lowest five level doublets could be calculated. Then we assumed that the transition to the isomer state $J^\pi = 6^-$ or low-spin, short-lived ground state $J^\pi = 1^-$ is defined by the "minimal multipolarity" rule. That means the states with spins $J > 3$ should populate the ground state, while those with $J \leq 3$ should feed the isomer state. Then the branching ratio is obtained as the ratio of the populations, derived from Eq. (7.12):

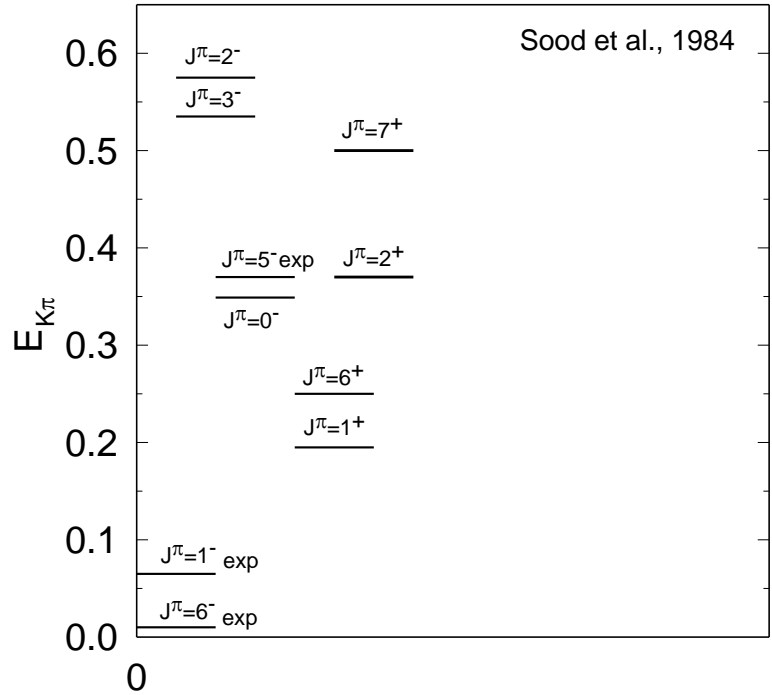

Fig. 7.16 Levels of $^{244}$Am.

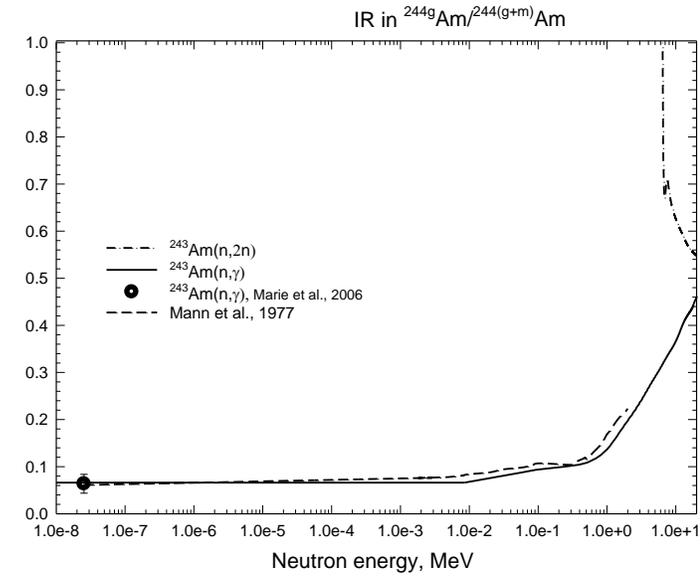

Fig. 7.17 Isomeric cross section ratio for the $^{243}$Am(n,γ)$^{242g}$Am and $^{243}$Am(n,γ)$^{242(m+g)}$Am reactions

$$r(E_n) = \frac{\displaystyle\sum_{J>(J_t+J_s)/2} W(U,J^\pi)}{\displaystyle\sum_{J\geq(J_t+J_s)/2} W(U,J^\pi)} \quad (7.12)$$

Figure 7.17 show the branching ratio, calculated for the modeled level schema, presented at Fig. 7.16.

The measured data reproduced assuming lihear extrapolation up to incident neutron energy 100 keV.

## 8. Fission cross section above emissive fission threshold

At incident neutron energies when fission of $^{243}$Am or $^{242}$Am nuclides is possible, as well as fission of $^{244}$Am, after emission of 1 or 2 pre-fission neutrons, the observed $^{243}$Am(n, F) fission cross section is a superposition of non-emissive or first chance fission of $^{244}$Am

$$\sigma_{nf}(E_n) = q(E_n)\frac{\pi \hat{\lambda}^2}{2(2I+1)}\sum_{lJ\pi}(2J+1)T_l(E_n)P_f^{J\pi}(E_n),\qquad(8.1)$$

and $x^{th}$-chance fission contributions as

$$\sigma_{nF}(E_n) = \sigma_{nf}(E_n) + \sum_{x=1}^{X}\sigma_{n,xnf}(E_n).\qquad(8.2)$$

The contributions to the observed fission cross section $\sigma_{n,xnf}(E_n)$, coming from (n, $x$nf), $x$= 1, 2, 3...X, fission of relevant equilibrated americium nuclei are weighted with a probability of $x$ neutron emission before fission. These cross sections are calculated as

$$\sigma_{n,xnf}(E_n) = \sum_{J\pi}\int_0^{U_{max}}W_{x+1}^{J\pi}(U)P_{f(x+1)}^{J\pi}(U)dU ,\qquad(8.3)$$

where $W_x^{J\pi}$ is the population of $(x+1)^{th}$ nucleus at excitation energy $U$ after emission of $x$ neutrons, excitation energy $U_{max}$ is defined by the incident neutron energy $E_n$ and the energy, removed from the composite system $^{244}$Am by the $^{243}$Am(n, $x$nf) pre-fission neutrons.

Contribution of first-chance fission $\sigma_{nf}(E_n)$ is defined by the pre-equilibrium emission of the first neutron and the fission probability $P_{f1}$ of the $^{242}$Am nuclide

$$\sigma_{f1} = \sigma_c(1-q(E))P_{f1}.\qquad(8.4)$$

Once the contribution of first neutron pre-equilibrium emission $q(E)$ is fixed, the first-chance fission probability $P_{f1}$ of the $^{244}$Am is defined by the level densities of fissioning $^{244}$Am and residual $^{243}$Am nuclides. Actually, it depends on the ratio of shell correction values $\delta W_{fA(B)}$ and $\delta W_n$. Different theoretical calculations of the shell corrections as well as of the fission barriers vary by 1-2 MeV. The same is true for the experimental shell corrections, which are obtained with a smooth component of potential energy calculated according to the liquid-drop or droplet model. However the isotopic changes of $\delta W_{fA(B)}$ and $\delta W_n$ are such that $P_{f1}$ viewed as a function of the difference $(\delta W_{fA(B)} - \delta W_n)$ is virtually independent on the choice of smooth component of potential energy. Therefore, we shall consider the adopted $\delta W_{fA(B)}$ estimates to be effective, provided that $\delta W_n$ are obtained with the liquid drop model. In the first "plateau" region and at higher energies we can safely use double-humped barrier model and relevant barrier parameters (see Table 8.1).

Consistent description of a most complete set of measured data on the (n, F), (n,2n), (n,3n) and (n,4n) reaction cross sections for the $^{238}$U target nuclide up to 20 MeV [64] enables one to consider the estimates of first neutron spectra emitted from the composite $^{244}$Am nuclide as fairly realistic. In case of $^{243}$Am(n, F) cross section, for which there are systematic discrepancies in measured data [12−27], calculated cross section is consistent with data by Knitter et al.[18], Fursov et al. [22], Fomushkin et al. [21, 23], Terayama et al. [17], Aiche et al. [20].

**Table 8.1** Fission barrier parameters of Am nuclei

| Nuclide | $E_{fA}$ | $\delta E_{fA}$ | sym.$_A$ | $E_{fB}$ | $\delta E_{fB}$ | sym.$_B$ | h$\omega_A$ | $\delta$h$\omega_A$ | h$\omega_B$ | $\delta$h$\omega_B$ | $\delta$ |
|---------|----------|-----------------|----------|----------|-----------------|----------|-------------|---------------------|-------------|---------------------|----------|
| $^{241}$Am | 6.3 | 0.5 | axial | 5.15 | 0.3 | mass-asym. | 0.8 | 0.2 | 0.52 | 0.2 | 0.0 |
| $^{242}$Am | 6.515 | 0.1 | Non-axial | 5.95 | 0.05 | mass-asym. | 0.7 | 0.05 | 0.44 | 0.2 | 0.02 |
| $^{243}$Am | 6.4 | 0.5 | axial | 5.00 | 0.3 | mass-asym. | 0.8 | 0.2 | 0.50 | 0.2 | 0.8 |
| $^{244}$Am | 6.35 | 0.1 | Non-axial | 5.92 | 0.05 | mass-asym. | 0.65 | 0.05 | 0.43 | 0.2 | 0.02 |

Figures 8.1-8.2 demonstrate the fission data fit from 100 keV up to 20 MeV. The contributions of emissive $^{243}$Am(n, nf) and $^{243}$Am(n, 2nf) fission to the total fission cross section are shown.

## 9. (n,2n) and (n,3n) cross section

The reaction chain $^{243}$Am(n,2n)$^{242m(g)}$Am was used to measure the yield of lower spin $^{242g}$Am state, by Norris-Gancartz (1982) [65] at ~15 MeV. From the ratio of the calculated yields at ~15 MeV it follows that the yield of low spin state is higher than that of high spin state. Near the threshold the ratio is defined by the relative position of states and feeding from the higher laying levels.

## 9.1 Branching ratio of short-lived $^{242g}$Am (1$^-$) and long-lived $^{242m}$Am (5$^-$) states in $^{243}$Am(n,2n) reaction

Ground $^{242g}$Am and isomer $^{242m}$Am states of the residual nuclide $^{242}$Am are excited in the reaction $^{243}$Am(n, 2n)$^{242g(m)}$Am. The $^{243}$Am(n,2n)$^{242g(m)}$Am reaction populates either the T$_{1/2}$ =16h ground state $^{242g}$Am with J$^π$=1$^-$ or the $^{242m}$Am isomer J$^π$=5$^-$ with T$_{1/2}$ =141y. The ground state $^{242g}$Am mostly β$^-$-decays to $^{242}$Cm, or goes to $^{242}$Pu via electron capture. The yield of the $^{242g}$Am short-lived ground state in the reaction chains $^{243}$Am(n,2n)$^{242g}$Am(β$^-$)$^{242}$Cm and $^{241}$Am(n, γ) $^{242g}$Am(β$^-$)$^{242}$Cm influences the α–activity and neutron activity of the spent fuel due to emerging nuclides $^{242}$Cm and $^{238}$Pu. The yield of the $^{242m}$Am long-lived isomer state, which due to large and odd value of J$^π$=5$^-$ may decay to $^{242g}$Am via isomeric transition IT only, emerging in reactions $^{243}$Am(n,2n)$^{242g}$Am(β$^-$)$^{242}$Cm and $^{241}$Am(n, γ)$^{242m}$Am influences the neutron activity of the spent fuel due to spontaneous fission of $^{242m}$Am. It gives a path for the $^{244}$Cm yield via $^{242m}$Am(n,γ)$^{243}$Am(n,γ)$^{244m}$Am(β$^-$(ε))$^{244}$Cm($^{244}$Pu) or $^{242m}$Am(n,γ)$^{243}$Am(n,γ)$^{244g}$Am(β$^-$)$^{244}$Cm. If not the forbidden β$^-$-decay of $^{242m}$Am state, the major path for the $^{244}$Cm build-up, would be closed.

The approach applied for the modeling ratio of the yields of short-lived (1$^-$) and long-lived (6$^-$) of $^{237}$Np(n,2n) $^{236s(l)}$Np reaction $r(E_n) = \sigma_{n2n}^l(E_n)/\sigma_{n2n}^s(E_n)$ from threshold energy up to 20 MeV allowed to infer the yields of the short-lived state $^{236s}$Np in $^{237}$Np(n,2n) reaction. The consistent description of the data base on cross sections [237]Np(n,F), [237]Np(n,2n)[236s]Np was achieved [5]. The branching ratio $r(E_n)$ was obtained by modeling the residual nuclide [236]Np levels. Excited levels of [236]Np are modeled using predicted Gallher-Moshkowski doublets.

In case of [243]Am(n,2n)[242m(g)]Am reaction the branching ratio $r(E_n) = \sigma_{n2n}^m(E_n) / \sigma_{n2n}^g(E_n)$ and are of interest. The branching ratio $r(E_n) = \sigma_{n2n}^m(E_n) / \sigma_{n2n}^g(E_n)$ from threshold energy to 20 MeV could be defined by the ratio of the populations of two lowest states in [242]Am [61]. These populations are defined by the γ-decay of the excited states, which is described by the kinetic equation, developed by Strutinsky et al. [62]. The branching ratio $r(E_n)$ is defined by the ratio of the populations of the two lowest states, [242g]Am, with spin $J = 1^-$ and [242m]Am, with spin $J = 5^-$.

The γ-decay of the excited nucleus described by the kinetic equation [46] as further developed by Dovbenko et al. [66]:

$$\frac{\partial \omega_k(U, J^\pi, t)}{\partial t} = \sum_{J'\pi'} \int_0^{U_g} \omega_{k-1}(U', J^{\pi'}, t) \frac{\Gamma_\gamma(U', J^{\pi'}, U, J^\pi)}{\Gamma(U', J^{\pi'})} dt - \omega_k(U, J^\pi, t) \frac{\Gamma_\gamma(U, J^\pi)}{\Gamma(U, J^\pi)}, \quad (9.1)$$

here $\omega_k(U, J^\pi, t)$ is the population of the state $J^\pi$ at excitation U at time t, after emission of $k$ γ-quanta; $\Gamma_\gamma(U', J^{\pi'}, U, J^\pi)$ is the partial width of γ-decay from the $(U', J^{\pi'})$ to the state $(U, J^\pi)$, while $\Gamma(U, J^\pi)$ is the total decay width of the state $(U, J^\pi)$. For any state $(U, J^\pi)$ with the excitation energy $0 \leq U \leq U_g$, the initial population is

$$\omega_k(U, J^\pi, t = 0) = \delta_{k0} \omega_0(U, J^\pi). \quad (9.2)$$

That equation means in the initial state we deal with the ensemble of states $(U, J^\pi)$, excited in [243]Am(n,2n) reaction. Integrating the Eq. (9.1) over $t$, one gets the population $W(U, J^\pi)$ of the state $(U, J^\pi)$ after emission of $k$ γ-quanta:

$$\omega_k(U, J^\pi, \infty) - \omega_k(U, J^\pi, 0) = \sum_{J'\pi'} \int_U^{U_g} \frac{\Gamma_\gamma(U', J^{\pi'}, U, J^\pi)}{\Gamma(U', J^{\pi'})} \int_0^\infty \omega_{k-1}(U', J^{\pi'}, t) dt dU' -$$
$$\frac{\Gamma_\gamma(U, J^\pi)}{\Gamma(U, J^\pi)} \int_0^\infty \omega_k(U, J^\pi, t) dt \quad (9.3)$$

Denoting the population of the state $(U, J^\pi)$ after emission of $k$ γ-quanta

$$W_k(U, J^\pi) = \frac{\Gamma_\gamma(U, J^\pi)}{\Gamma(U, J^\pi)} \int_0^\infty \omega_k(U, J^\pi, t) dt, \quad (9.4)$$

and taking into account the condition that $\omega_k(U, J^\pi, \infty) = 0$ for any state, belonging to ensemble $(U, J^\pi)$, Eq. (9.3) could be rewritten as

$$W_k(U, J^\pi) = \sum_{J'\pi'} \int_U^{U_g} \frac{\Gamma_\gamma(U', J^{\pi'}, U, J^\pi)}{\Gamma(U', J^{\pi'})} W_{k-1}(U', J^{\pi'}) dU' + \omega_k(U, J^\pi, 0). \quad (9.5)$$

The population of any state $(U, J^\pi)$ after emission of any number of γ-quanta is a lumped sum

$$W(U, J^\pi) = \sum_k W_k(U, J^\pi), \quad (9.6)$$

then from Eq. (9.5) one easily gets

$$W(U,J^{\pi}) = \sum_{J'\pi'} \int_{U}^{U_g} \frac{\Gamma_{\gamma}(U',J^{\pi'},U,J^{\pi})}{\Gamma(U',J^{\pi'})} W(U',J^{\pi'})dU' + W_0(U,J^{\pi}).$$  (9.7)

The integral equation (9.7) in the code STAPRE [63] is solved as a system of linear equations, the integration range $(U, U_g)$ is binned, in the assumption that there are no γ-transitions inside the bins.

The isomer branching ratio depends mostly on the low-lying levels scheme and relevant γ-transitions probabilities. Though experimental data are available for $^{242}$Am [32], we will use a simplified approach, since experimental level scheme and gamma-decay intensities are still incomplete. Modeling of low-lying levels of $^{242}$Am in [61] is accomplished based on the assumption that ground and first few excited states are of two-quasi-particle nature. For actinides with quadrupole deformations the superposition principle is usually adopted, the band-head energies of the doubly-odd nucleus are generated by adding to the each unpaired configuration $(\Omega_p, \Omega_n)$, as observed in the isotopic/isotonic (A-1) nucleus, the rotational energy contribution and residual (n-p) interaction energy contribution. The angular momenta of neutron and proton quasi-particles could be parallel or anti-parallel. In the independent quasi-particle model the two-quasi-particle states, $K^+ = |K_n + K_p|$ and $K^- = |K_n - K_p|$, are degenerate. Gallaher-Moshkowski doublets [61] appear because of (n-p) residual interaction. Fig 9.1 shows employed band-head energies for the two-quasi-particle states expected in the odd-odd nuclide $^{242}$Am up to 700 keV. The spectroscopic properties of two pairs of proton and neutron single particle states were derived from those experimentally observed in the isotopic (Z=95) and isotonic (N=147) odd-mass nuclei with mass (A-1). Figure 9.1 shows levels expected, which have similar ordering as experimentally observed [32]. For the band-heads, shown on Figure 9.1, the rotational bands were generated as

$$E_{JK\pi} = E_{JK} + 5.5\left[J(J+1) - K(K+1)\right].$$  (9.8)

Obviously, the schema presented on Fig. 9.1 does not represent a complete set to allow the calculation of absolute yields of $^{242}$Am (n,γ) $^{242m}$Am and $^{242}$Am (n,γ) $^{242g}$Am reactions. Rotational bands were constructed up to 700 keV excitation, modeling levels with spins $J^{\pi} \leq 10$, in total up to 70 levels. The simple estimate [29] of the number of levels in odd-odd nuclei as

$$N(U) = e^{2\Delta_0/T}(e^{U/T} - 1),$$  (9.9)

predicts up to 280 level at $U\sim700$ keV, $T$=0.388 MeV, $\Delta$=12/$A^{1/2}$, MeV. We assume that the modeled angular momentum distribution would not be much different from realistic estimates. Since the complete data on the γ-transitions are missing, we assumed the simple decay scheme: only E1, E2 and M1 transitions are allowed in a continuum excitation energy range. Inter-band transitions forbidden, i.e., only γ-transitions within the rotational bands are possible. In that approach the populations of the lowest five level doublets could be calculated. Then we assumed that the transition to the isomer state $J^{\pi} = 5^-$ or low-spin, short-lived ground state $J^{\pi} = 1^-$ is defined by the "minimal multipolarity" rule. That means the states with spins $J < 3$ should populate the ground state, while those with $J \leq 3$ should feed the isomer state. Then the branching ratio could be obtained as the ratio of the populations, derived from Eq. (9.7):

$$r(E_n) = \frac{\sum_{J<(J_I+J_s)/2} W(U,J^{\pi})}{\sum_{J\geq(J_I+J_s)/2} W(U,J^{\pi})}$$  (9.10)

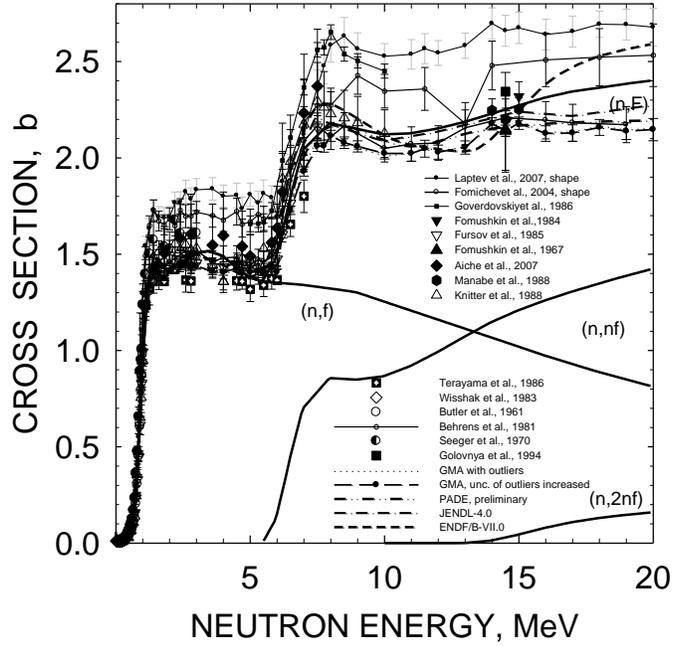

Fig. 8.1 Fission cross section of $^{243}$Am(n, f).

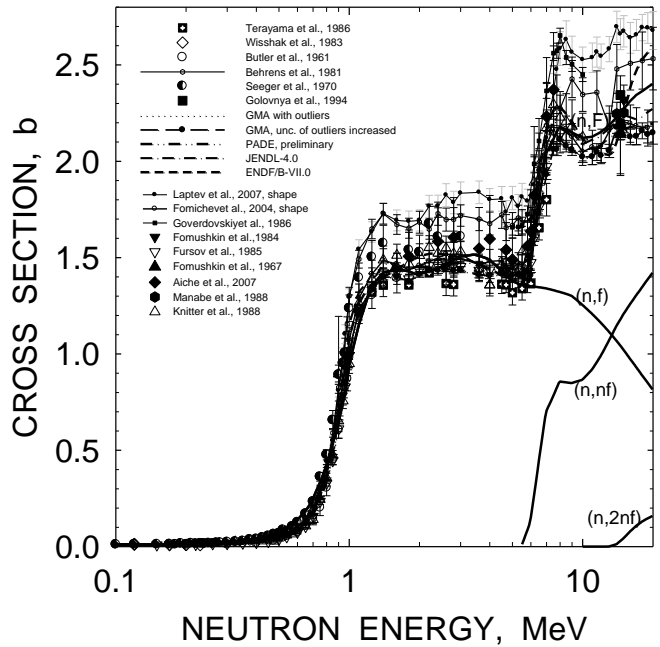

Fig. 8.2 Fission cross section of $^{243}$Am(n, f).

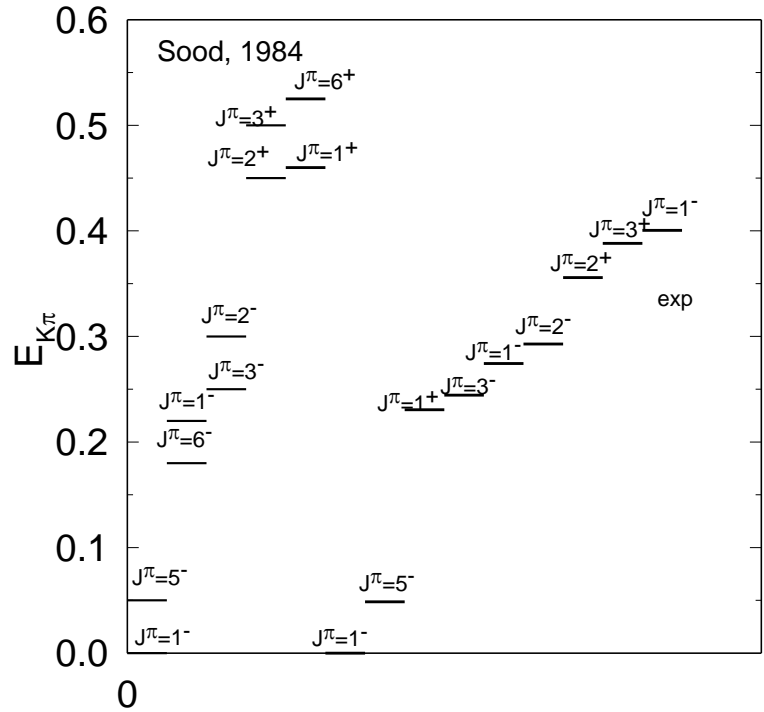

Fig. 9.1 Levels of $^{242}$Am.

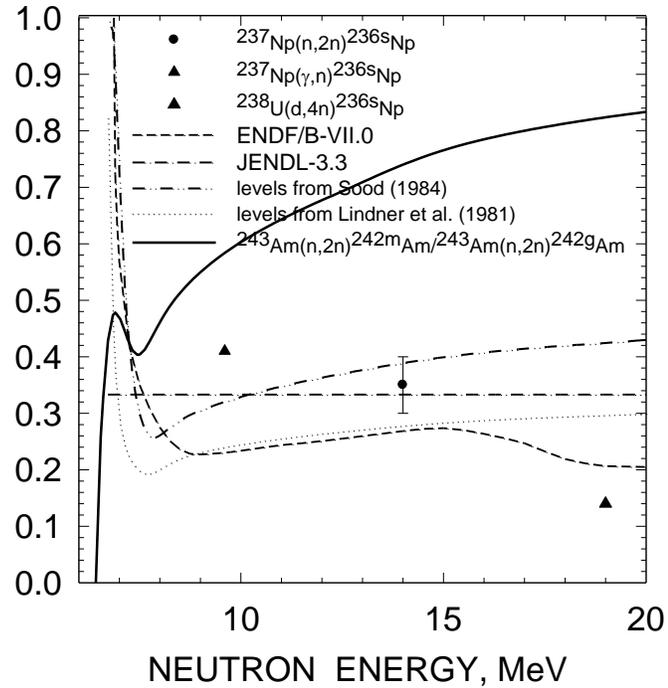

Fig. 9.2 Branching ratio of $^{243}$Am(n,2n)$^{242m}$Am/$^{243}$Am(n,2n)$^{242g}$Am reaction

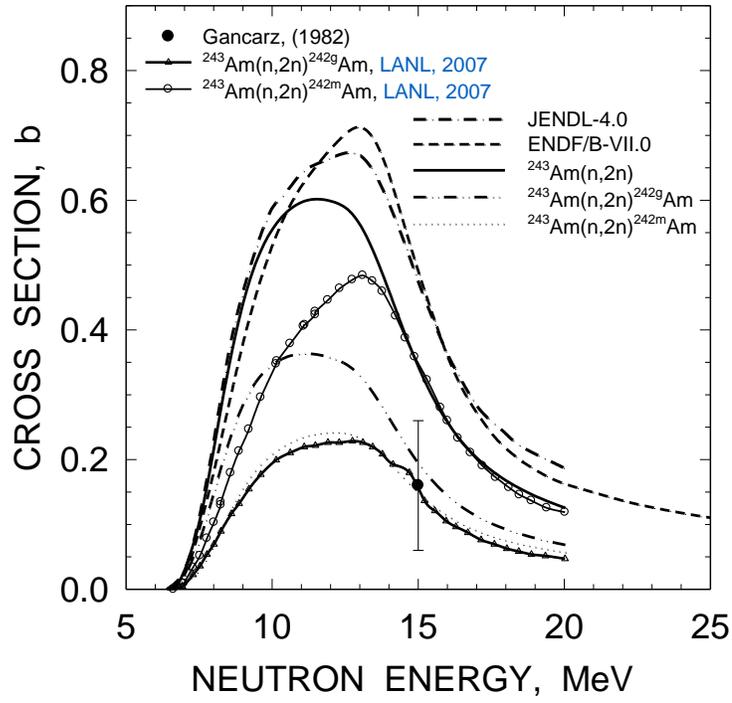

Fig. 9.3 Cross section of $^{243}$Am (n, 2n) reaction

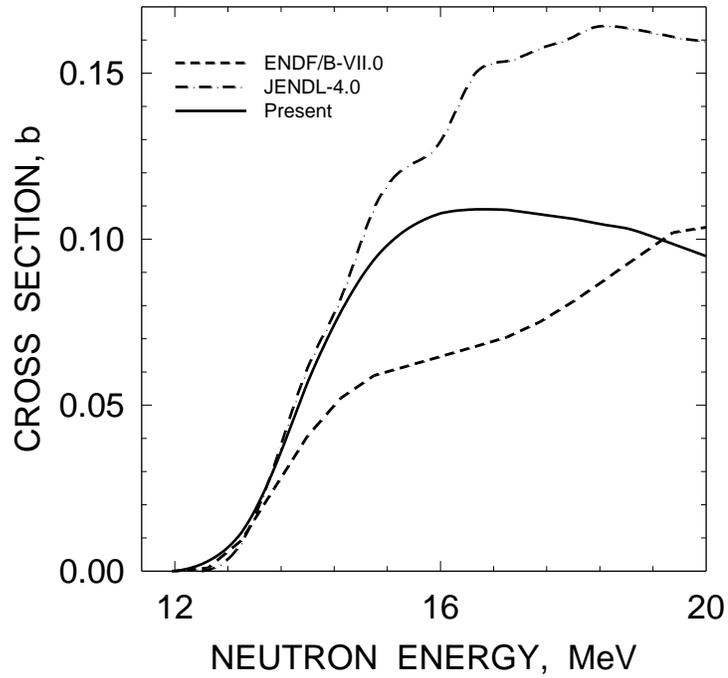

Fig. 9.4 Cross section of $^{243}$Am (n, 3n) reaction

Figure 9.2 show the branching ratio, calculated for the modeled level schema, presented at Fig. 9.1. The modeled level scheme appears to be quite compatible with the measured data on the $^{242g}$Am ground state yield [65].

Figure 9.2 shows, that the yields of the $^{242g}$Am and $^{242m}$Am at ~15 MeV are still comparable, the latter being lower, as it should be for higher spin state in (n,2n) reaction. In calculation of [5] the ratio is opposite different. The branching ratio for $^{237}$Np(n,2n) reaction shown on Fig. 9.2 is much different from that calculated here for $^{243}$Am(n,2n)$^{242m(g)}$Am reaction. It is explained by the level spectra differences for residual nuclei. Figure 9.3 shows respective cross sections for the $^{243}$Am(n,2n)$^{242m(g)}$Am reaction. Calculated $^{243}$Am(n,3n) cross section is shown on Fig. 9.3. It should be stressed once again, that we base our evaluation of fission cross section on simultaneous statistical model description of both (n,2n) and (n, F) reactions for $^{241,243}$Am.

# 10. Evaluation of prompt neutron yield in $^{243}$Am(n, F)

Average prompt neutron yield in neutron−induced fission of $^{243}$Am target nuclide was evaluated in a non-model least−squares fit of the experimental data with the use of the GMA code. The experimental data obtained by two groups. Data by Khokhlov et al. [67], measured relative $^{252}$Cf(sf) prompt neutron yield, were renormalized to the latest evaluation recommended as standard [10]. Data by Frehaut et al. [68], which are not available in EXFOR data base, were digitized from the figure given in the [68]. They are in good agreement with data [67] in the overlapping energy region. Non-model least squares fit done with the GMA code [9, 10]. Chi−squared value per degree of freedom is about 1. For final presentation of the evaluation, 2-nd order polynomial fit was done for the GMA evaluated values below ~15 MeV and data of model calculations [64] above ~15 MeV. The thermal value adopted equal to value stemming from the fit at 0.1 keV, taking into account common for most nuclides energy dependence of prompt fission neutron yields in the resonance region [10].

The evaluated data are shown at Fig. 10.1 in comparison with experimental data, non-model fit and values from various evaluated data libraries. The uncertainties of evaluated data vary in-between 0.6% and 2% in the range where experimental data are available. The thermal value of $<\nu_d>$ is evaluated as 3.155±0.035, it is lower than that of ENDF/B-VII.0 [30] evaluation but consistent with JENDL-4.0 value.

We applied more complicated theoretical approach, than just linear extrapolation of $\nu_p$, with incorporation of the pre-fission neutron emission and contributing fission chances. At incident neutron energies above emissive fission threshold the number of prompt fission neutron $\nu_p$ calculated as

$$\nu(E_n) = \sum_{x=1}^{X} \left(\nu_x(E_{nx}) + (x-1)\right)\cdot\beta_x(E_n),\qquad(10.1)$$

here $x$ =1, …$X$ is the multiplicity for the $x$−chance fission of the nuclei A+1, A, A-1, A-2 after emission of $(x-1)$ pre-fission neutrons, $\beta_x(E_n)$ is the $x$ −chance contribution to the observed fission cross section, $\nu_x(E_{nx})$ - prompt neutron multiplicity for x-th fissioning nucleus. The excitation energy of A, A-1, A-2 nuclides, which emerge after emission of (x-1) pre-fission neutrons is defined

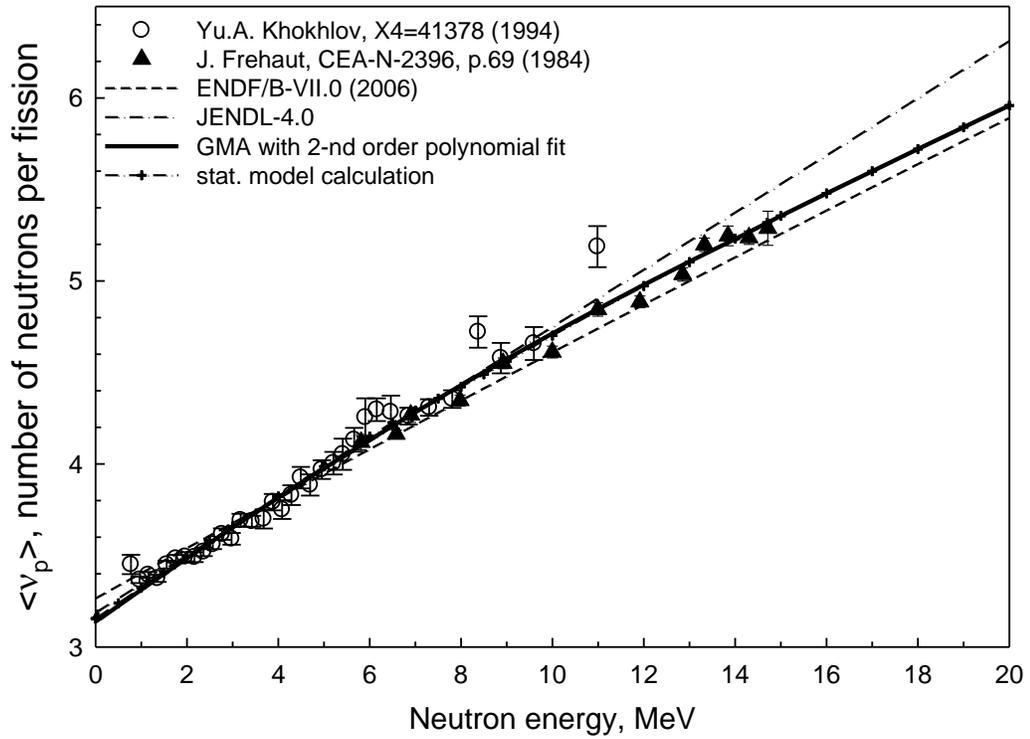

Fig. 10.1 Prompt fission neutron number of $^{243}$Am

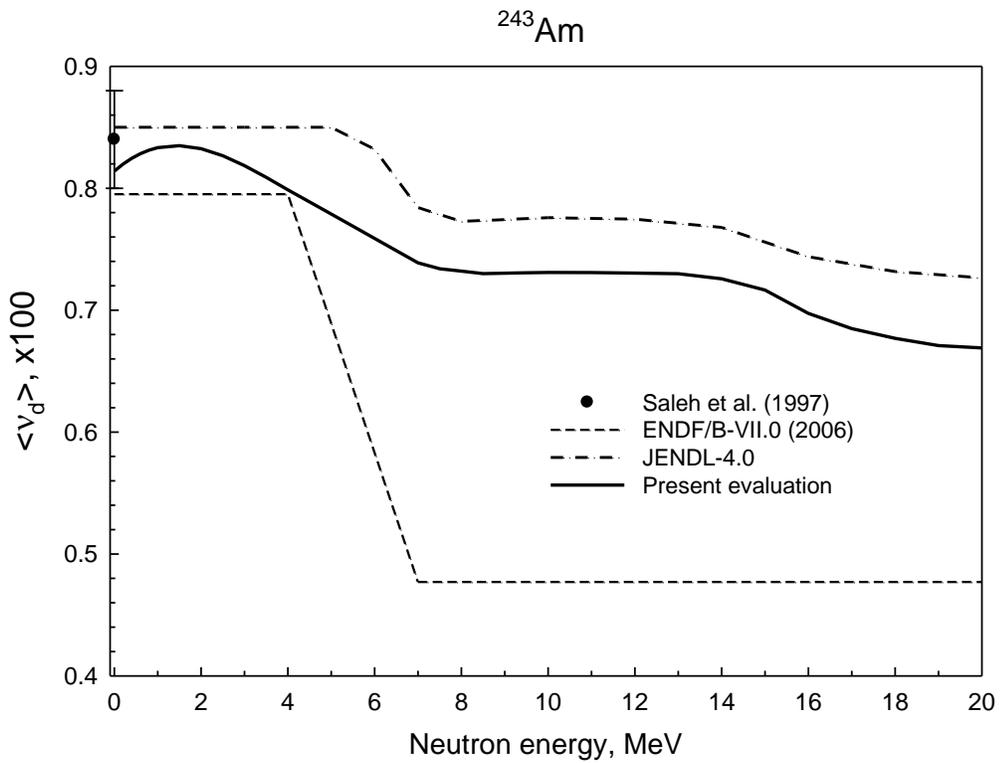

Fig. 11.1 Delayed fission neutron number of $^{243}$Am

as

$$E_{nx} = E_n - \sum_j B_{nj} - \left\langle E_{xj} \right\rangle, \qquad (10.2)$$

here $B_{nj}$ - neutron binding energy for the (A+1-j), j=1, 2, 3, nucleus, $\left\langle E_{xj} \right\rangle$ - average energy of j-th pre-fission neutron. The incident neutron energy dependence of neutron multiplicity in the energy range $E_n \leq 6$ MeV for all fissioning $^{244, 243,242,241}$Am nuclei was taken from evaluation by Malinovskij [69], to reproduce the measured data on $\nu_p$ (see Table 10.1). We assumed that excitation energy $E_{nx}$ is brought into $A_j$ nuclide with the reaction: n+( $A_j$ -1) →fission. Incident neutron energy in this hypothetical reaction equals to $(E_{nj} - B_{nAj})$. In this way the $\nu_x(E_{nx})$ functions for all nuclides in the mass chain $^{242,241,240,239}$Am were calculated. Energy dependence of $\nu_p$ versus incident neutron energy estimated with this equation is compared on Fig. 10.2 with previous present GMA-evaluation and previous evaluations. Relevant partial contributions to $\nu_p$ are shown on Fig. 10.1. "Step" in $\nu_p$ around (n,nf) reaction threshold is due to the pre-fission neutrons, emitted in $^{243}$Am(n,nf) reaction. The similar behavior was evidenced in measured data for $^{232}$Th(n, F) and $^{238}$U(n,f), it was reproduced with the present model [64].

**Table 10.1** Evaluated [64] first chance $\nu_p$ –values for $^{243,242,241,240}$Am target nuclides

| Target | $\nu_p^{th}$ | $\nu_p$($E_n$ MeV) | $\nu_p$(6 MeV) |
|--------|--------------|--------------------|----------------|
| $^{243}$Am | 3.260* | 3.709 (3.0) | 4.150 |
| $^{242}$Am | 3.257 | 3.696 (3.0) | 4.127 |
| $^{241}$Am | 3.090* | 3.624 (3.0) | 4.047 |
| $^{240}$Am | 3.205 | 3.623 (3.0) | 4.034 |

\*)tuned to measured data

# 11. Evaluation of averaged delayed neutron yields for $^{243}$Am (n, F)

There is only one result of direct measurements of total average delayed neutron yield for $^{243}$Am(n, f) - measurements by Saleh et al. [70] for thermal neutron spectrum. In these conditions, the results of evaluation given in ENDF/B-VII.0 [30] and JENDL-4 [13] and shown at Fig. 11.1 are rather discrepant. JENDL-4 evaluation obtained in calculations with account of contribution from (n, f), (n, nf) and (n, 2nf) chances of the observed fission, considering accordingly the evaluated data available for $^{243}$Am(n, f) below the threshold of (n, n'f) reaction, $^{242m}$Am(n, f) and $^{241}$Am(n, f). The energy dependence of the delayed neutron yield for the first chance fission was assumed negligible and the respective yield has a constant value for excitations below the threshold of (n, n'f) reaction. In present evaluation, the energy dependence of average delayed neutron yield was modeled comparing the values of the group yields for $^{243}$Am and $^{237}$Np at thermal energy and relative energy dependence of the group yields known for $^{237}$Np [71]. The results of the final evaluation are shown in Fig. 11.1.

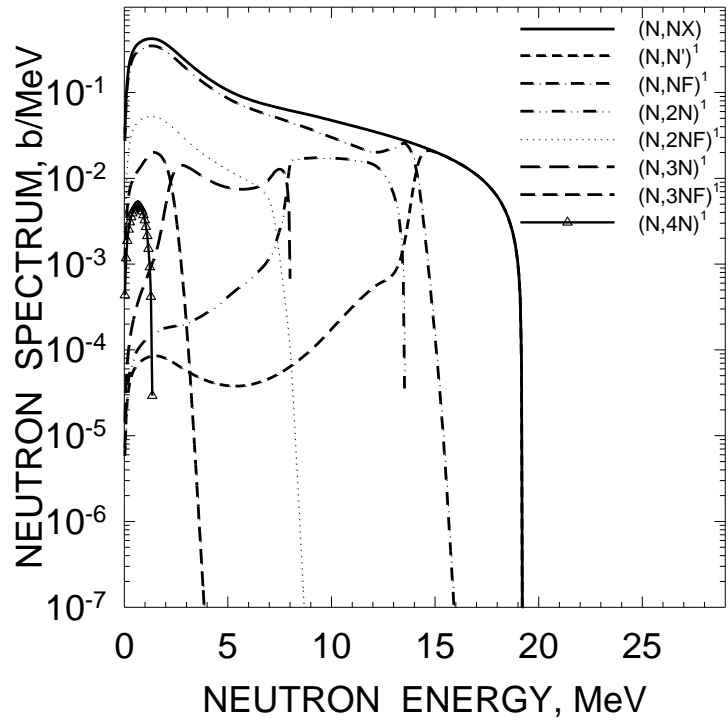

Fig. 12.1 Components of first neutron spectrum of $^{243}$Am +n interaction

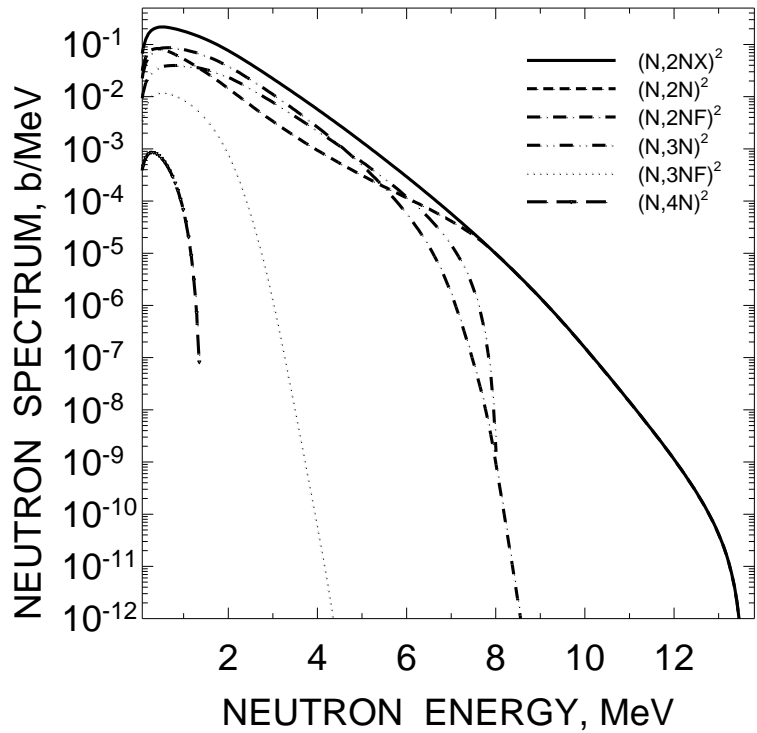

for incident neutron energy 20 MeV.

Fig. 12.2 Components of second neutron spectrum of $^{243}$Am +n interaction for incident neutron energy 20 MeV.

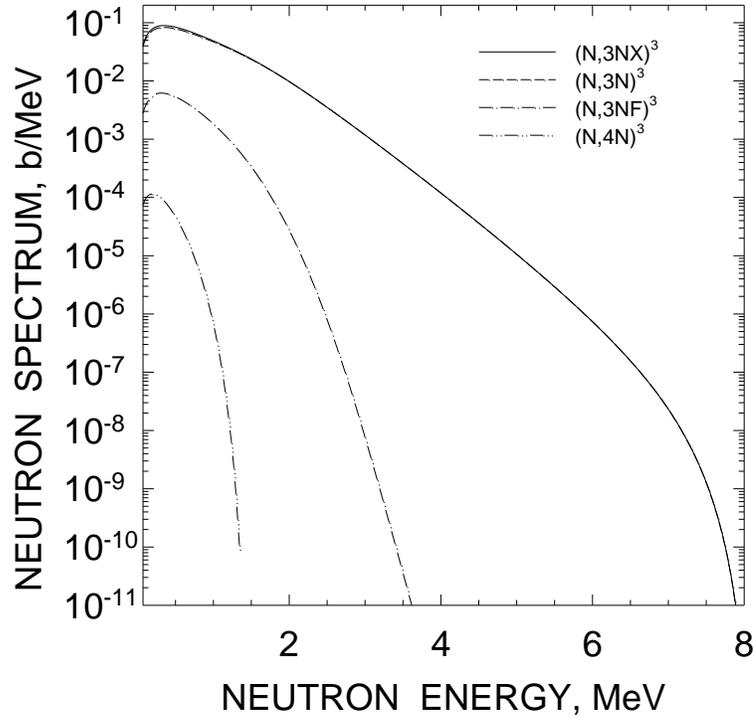

Fig. 12.3 Components of third neutron spectrum of $^{243}$Am +n interaction for incident neutron energy 20 MeV.

## 12. Energy distributions of secondary neutrons

Energy distributions for (n, 2nγ), (n,3nγ) and (n, n'γ) reactions were calculated with a Hauser-Feshbach statistical model of cascade neutron emission [5, 64], taking into account exclusive pre-fission (n, xnf) and (n, xnγ) neutron spectra, with the allowance of pre-equilibrium emission of first neutron.

Prompt fission neutron spectra (PFNS) calculated with a phenomenological model, developed for the first-chance fission by Kornilov et al. [72]. The model extended to the emissive fission domain for various targets by Maslov et al. [5, 64, 73] and exclusive prefission neutron spectra of $^{243}$Am(n, xnf) and $^{243}$Am(n, xnγ) being included. Exclusive pre-fission neutron spectra of (n,xnf) reactions, either equilibrium and pre-equilibrium spectra of pre-fission (n,xnf) neutrons are strictly correlated with (n, F) and (n, xn) reaction cross sections. This approach was used previously for the description of the PFNS and neutron emission spectra for $^{238}$U+n [64], $^{235}$U+n [73], $^{232}$Th+n [64] and $^{237}$Np+n [5] interactions. A number of experimental signatures were revealed and correlated with the exclusive pre-fission (n, xnf) and (n, xnγ) neutron spectra. This validated approach is used for the $^{243}$Am(n, F), $^{243}$Am(n, 2nγ) and PFNS description/prediction for n+$^{243}$Am interaction for non-emissive and emissive fission domain. Average energies of PFNS would predict distinct lowering in the vicinity of (n, nf) and (n, 2nf) reaction thresholds well known in measured PFNS shapes for major actinides.

### 12.1 (n,xnγ) and (n,xnf) neutron emission spectra

Exclusive (n, xnγ) and (n,xnf) neutron emission spectra for x = 1, 2, 3, reactions are calculated with Hauser-Feshbach model taking into account fission and gamma-emission competition to neutron emission, actually neutron spectra are calculated simultaneously with fission and (n,xn) reaction cross sections. The pre-equilibrium emission of first neutron is fixed by the description of high energy tails of (n,2nγ) reaction cross sections and (n, F) reaction cross sections for $^{237}$Np [5], $^{235}$U [73], $^{238}$U and $^{232}$Th target nuclides [64].

First neutron spectrum of the $^{241}$Am(n, nf) or $^{241}$Am(n, nγ) reactions is the sum of evaporated and pre-equilibrium emitted neutron contributions. Second and third neutron spectra for $^{243}$Am(n, xnf) and $^{243}$Am(n, xnγ) reactions are assumed to be evaporative. Pre-fission neutron spectrum of $^{243}$Am(n, nf) reaction, especially its hard energy tail, is sensitive to the description of fission probability of $^{243}$Am nuclide near fission threshold.

Components of first neutron spectra for $E_n = 20$ MeV are shown on Fig. 12.1. Components of second neutron spectra for $E_n = 20$MeV are shown on Fig. 12.2. Components of third neutron spectra for $E_n = 20$ MeV are shown on Fig. 12.3.

That is an illustration of the strong dependence of the partial contributions of the exclusive first neutron spectra on the fissilities of the composite (A+1) nuclides as well as relative fissilities of A, A-1, A-2 nuclides.

Summarizing, we anticipate that partial (n, xnf) pre-fission neutron spectra for $^{243}$Am target nuclide would be pronounced in observed PFNS to a different extent as compared with $^{238}$U+n [64], $^{235}$U+n [73], $^{232}$Th+n [64] and $^{237}$Np+n [5] interactions.

## 12.2 Prompt Fission Neutron Spectra

PFNS from fission fragments calculated as a superposition of two Watt [74] distributions for heavy and light fission fragments (FF), the partial contributions being equal, while the temperatures of the fragments are different [72]. Fission fragments' kinetic energy is the superimposed phenomenological parameter, generally lower, than total kinetic energy (TKE) of accelerated fission fragments.

The prompt fission neutron spectrum $S\ (\varepsilon, E_n)$ is calculated as a sum of two Watt distributions, modified to take into account the emission of prompt fission neutrons before full acceleration of fission fragments. The neutrons, emitted from heavy and light fission fragments are included with equal weights:

$$S(\varepsilon, E_n) = 0.5 \cdot \sum_{j=l,h} W_j(\varepsilon, E_n, T_j(E_n), \alpha), \tag{12.1}$$

$$W_j(\varepsilon, E_n, T_j(E_n), \alpha) = \frac{2}{\sqrt{\pi} T_j^{3/2}} \sqrt{\varepsilon} \exp\left(-\frac{\varepsilon}{T_j}\right) \exp\left(-\frac{E_{vj}}{T_j}\right) \frac{sh(\sqrt{b_j \varepsilon})}{\sqrt{b_j \varepsilon}}, \tag{12.2}$$

$$b_j = \frac{4E_{vj}^0}{T_j^2}, \ T_j = k_j \sqrt{E_j^*} = k_j \sqrt{E_r - TKE + E_n + B_n}, \tag{12.3}$$

$$E_{vl}^0 = \frac{A_h}{A_l A} \cdot \alpha \cdot TKE, \ E_{vh}^0 = \frac{A_l}{A_h A} \cdot \alpha \cdot TKE. \tag{12.4}$$

The coefficient α is the ratio of the kinetic energies of the fragments at the moment of neutron emission to the kinetic energy of fully accelerated fragments and is, in fact, a free parameter. The ratio of the temperatures of the light and heavy fragment $r=T_l/T_h$ is another free parameter, which ensures the model [64] flexibility to reproduce the soft and hard tails of the PFNS. The parameters $\alpha$=0.808 and $r$=1.248 were fixed in [5] by fitting, in fact, the PFNS for n+$^{237}$Np system at $E_n$=7.8 MeV. For n+$^{243}$Am they are defined by systematics [72].

Exclusive (n,xnf) pre-fission neutron spectra, as described above, are calculated. At $E_n$ higher than the emissive fission threshold $S$ $(\varepsilon,E_n)$ is calculated as a superposition of pre-fission (n,xnf) neutrons $-d\sigma_{nxnf}^k/d\varepsilon$ (x=1, 2, 3, 4; k=1,…,x) and post-fission spectra $S_{A+2-x}(\varepsilon,E_n)$ of the neutrons from the fission fragments:

$$S(\varepsilon,E_n) = \widetilde{S}_{A+1}(\varepsilon,E_n) + \widetilde{S}_A(\varepsilon,E_n) + \widetilde{S}_{A-1}(\varepsilon,E_n) + \widetilde{S}_{A-2}(\varepsilon,E_n) =$$

$$\nu_p^{-1}(E_n) \cdot \big\{ \nu_{p1}(E_n) \cdot \beta_1(E_n)S_{A+1}(\varepsilon,E_n) + \nu_{p2}(E_n - \langle E_{nnf} \rangle)\beta_2(E_n)S_A(\varepsilon,E_n) +$$

$$+ \beta_2(E_n)\frac{d\sigma_{nnf}^1(\varepsilon,E_n)}{d\varepsilon} + \nu_{p3}(E_n - B_n^A - \langle E_{n2nf}^1 \rangle - \langle E_{n2nf}^2 \rangle)\beta_3(E_n)S_{A-1}(\varepsilon,E_n) + \beta_3(E_n) \cdot \qquad (12.5)$$

$$\left[\frac{d\sigma_{n2nf}^1(\varepsilon,E_n)}{d\varepsilon} + \frac{d\sigma_{n2nf}^2(\varepsilon,E_n)}{d\varepsilon}\right] + \nu_{p4}(E_n - B_n^A - B_n^{A-1} - \langle E_{n3nf}^1 \rangle - \langle E_{n3nf}^2 \rangle - \langle E_{n3nf}^3 \rangle) \cdot$$

$$\beta_4(E_n)S_{A-2}(\varepsilon,E_n) + \beta_4(E_n)\left[\frac{d\sigma_{n3nf}^1(\varepsilon,E_n)}{d\varepsilon} + \frac{d\sigma_{n3nf}^2(\varepsilon,E_n)}{d\varepsilon} + \frac{d\sigma_{n2nf}^3(\varepsilon,E_n)}{d\varepsilon}\right] \big\}.$$

Figures 12.4-12.7 compare present PFNS with those of JENDL-4.0 [13]. Some shape differences for the first- and higher chance fission noticed. Aerage energies (Fig. 12.8) of emitted prompt fission neutron spectra both in first-chance and emissive fission domains are discrepant.

Figure 12.9 shows the partial contributions of $^{243}$Am (n, f) and $^{243}$Am(n, xnf) reactions to the observed PFNS, shown on previous Fig. 12.7. The contribution of $^{243}$Am(n, nf) reaction in the soft part of the spectrum is systematically lower than that of $^{243}$Am(n, f) reaction. The contribution of (n,nf) reaction is higher in case of JENDL-4.0 [13] for the hard part of the spectrum.

The combined effect of fission chances and exclusive pre-fission neutron spectra leads to the lowering of the average energy of the PFNS of $^{243}$Am (n, F) in the vicinity of the $^{243}$Am (n, nf) and $^{243}$Am (n, 2nf) reaction thresholds. The dips are evidenced in measured PFNS average energy $\langle E \rangle$ of different U nuclei. In JENDL-4.0 [13] the PFNS is calculated with Madland-Nix model [75], the pre-fission neutron spectra are calculated, presumably, as exclusive ones. However, some discrepancies are noticed with measured data for $^{237}$Np(n, F), a second dip due to (n, 2nf) reaction is not reproduced in JENDL-4.0 [13] both for $^{237}$Np and $^{243,241}$Am.

## 12.3 (n,xn) Reactions Neutron Spectra

There is no measured data on neutron emission spectra for $^{243}$Am +n interaction. For incident neutron energy higher than emissive fission threshold, emissive neutron spectra are de-convoluted, components of 1st, 2nd and 3d neutron spectra are provided, where applicable. We have calculated 1st, 2nd and 3d neutron spectra for the (n, n$\gamma$), (n, 2n) and (n, 3n) reactions.

According to the ENDF/B-VI format specifications the secondary neutron spectra were summed up and tabular spectra for the (n,n$\gamma$), (n,2n) and (n,3n) reactions were obtained.

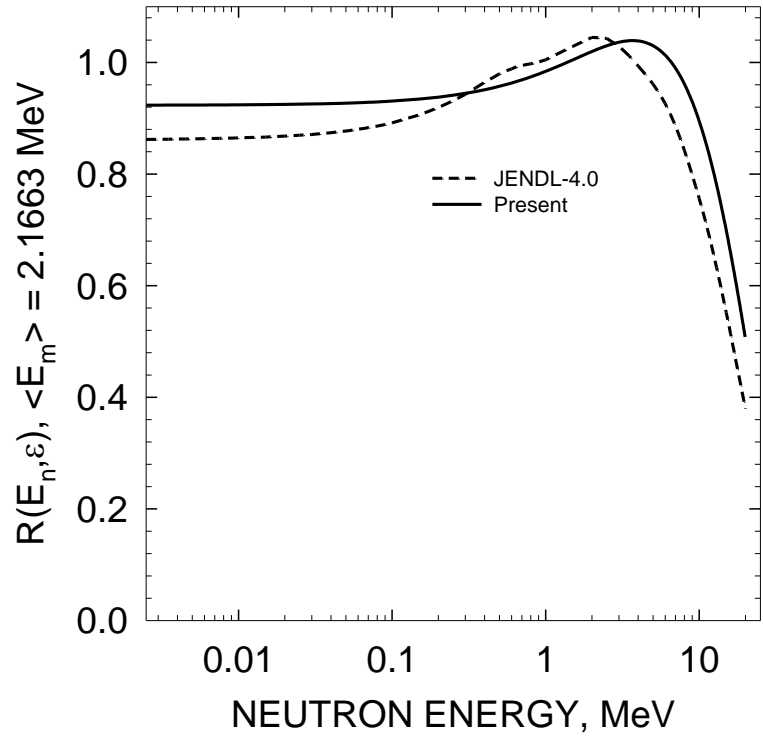

Fig. 12.4 Comparison of the prompt fission neutron spectrum for $^{243}$Am (n, F) reaction at incident neutron energy of $10^{-5}$ MeV.

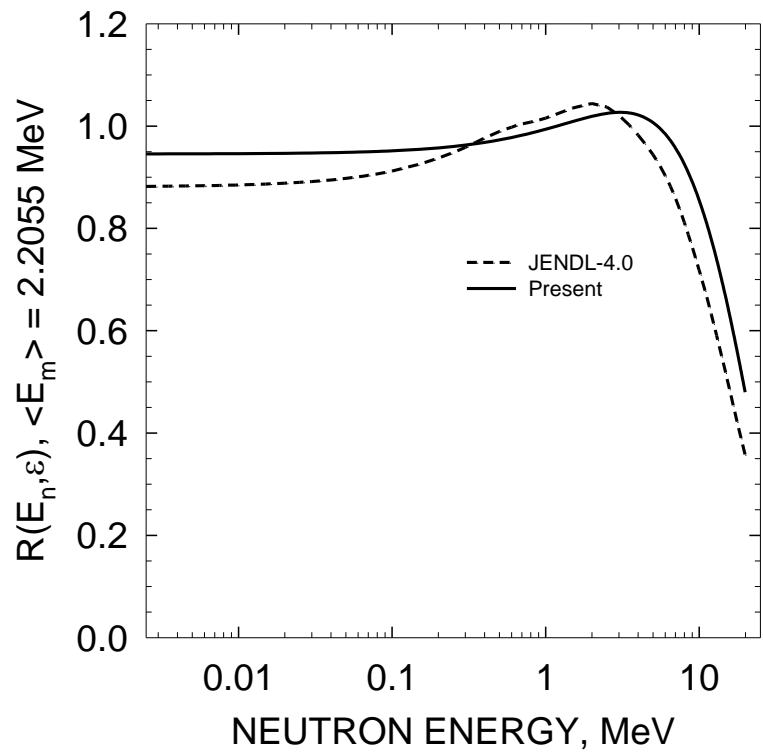

Fig. 12.5 Comparison of the prompt fission neutron spectrum for $^{243}$Am (n, F) reaction atincident neutron energy of 2 MeV.

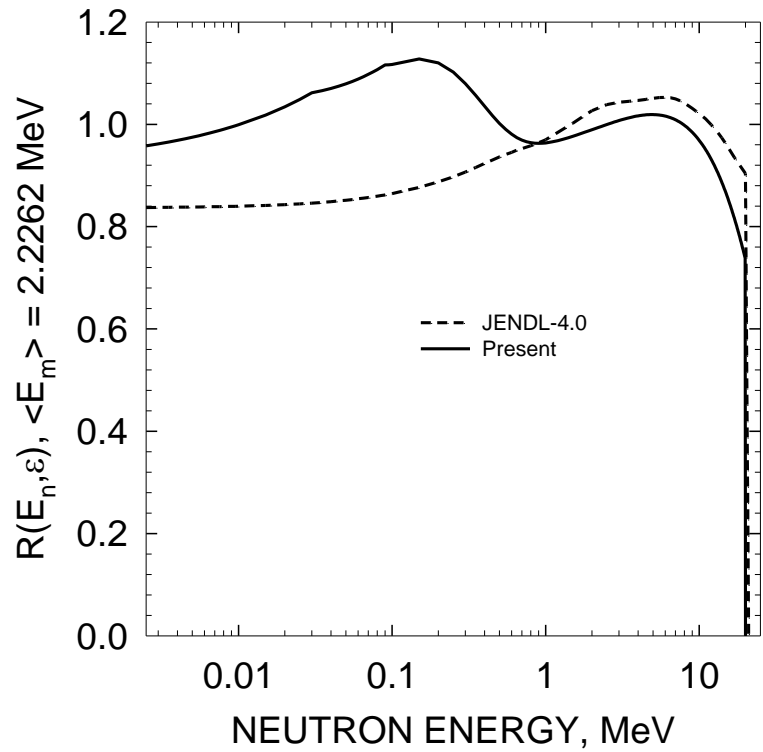

Fig. 12.6 Comparison of the prompt fission neutron spectrum for $^{243}$Am (n, F) reaction at incident neutron energy of 6 MeV.

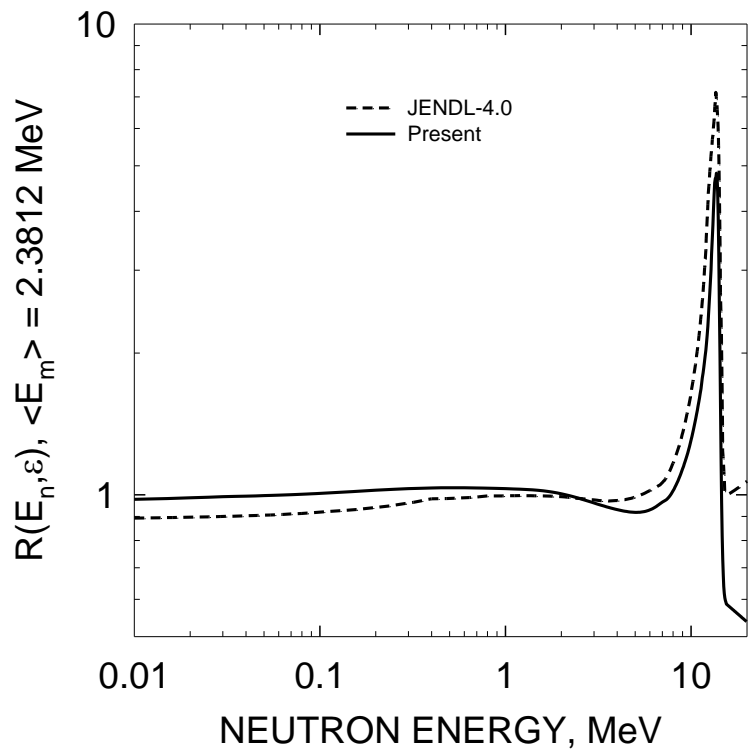

Fig. 12.7 Comparison of the prompt fission neutron spectrum for $^{243}$Am (n, F) reaction at incident neutron energy of 20 MeV.

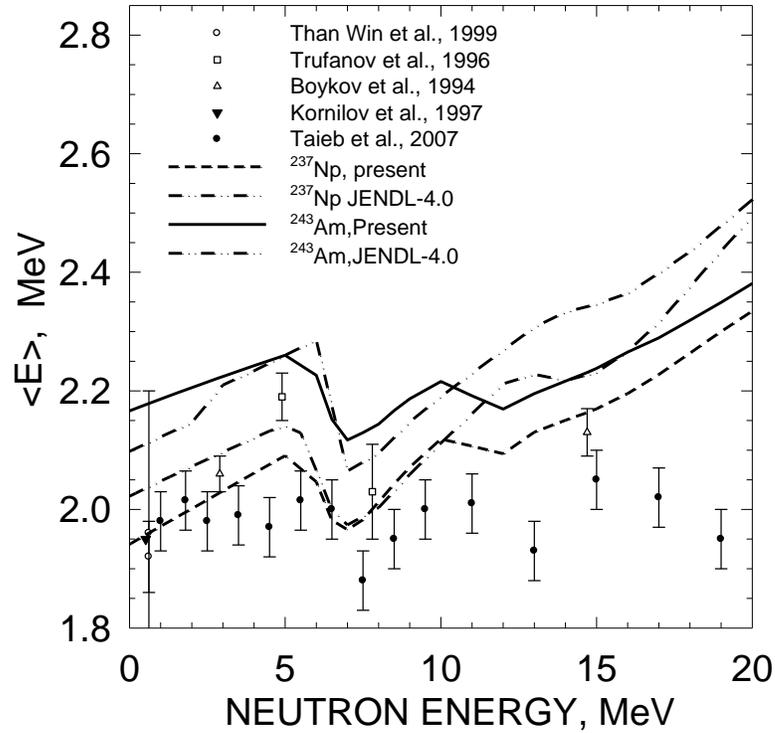

Fig. 12.8 Dependence of average energy of $^{243}$Am (n, F) and $^{237}$Np(n,F) prompt fission neutron spectra on the incident neutron energy.

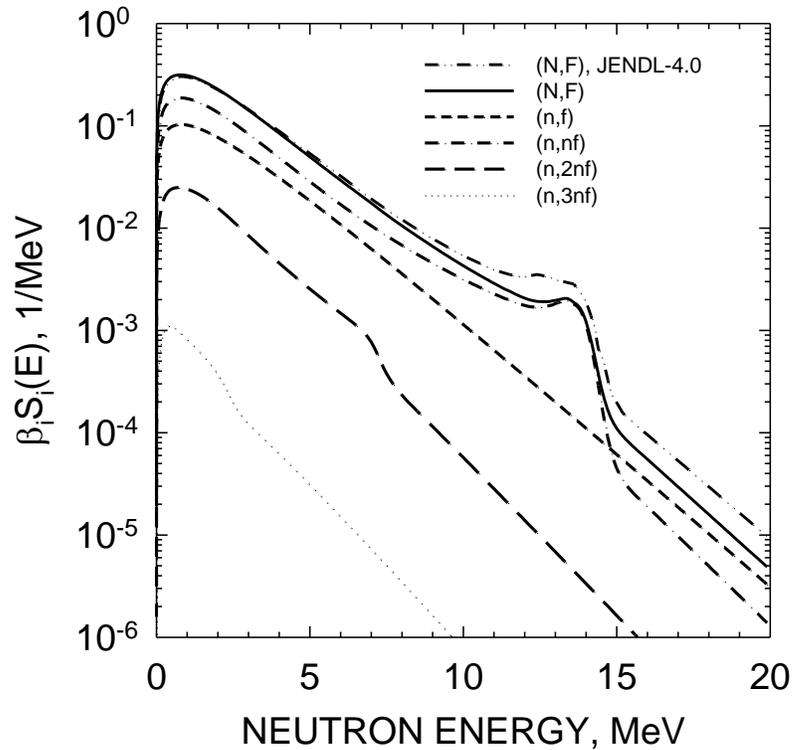

Fig. 12.9 Multiple-chance fission contributions to the prompt fission neutron spectrum for $^{243}$Am (n, F) reaction, incident neutron energy 20 MeV.

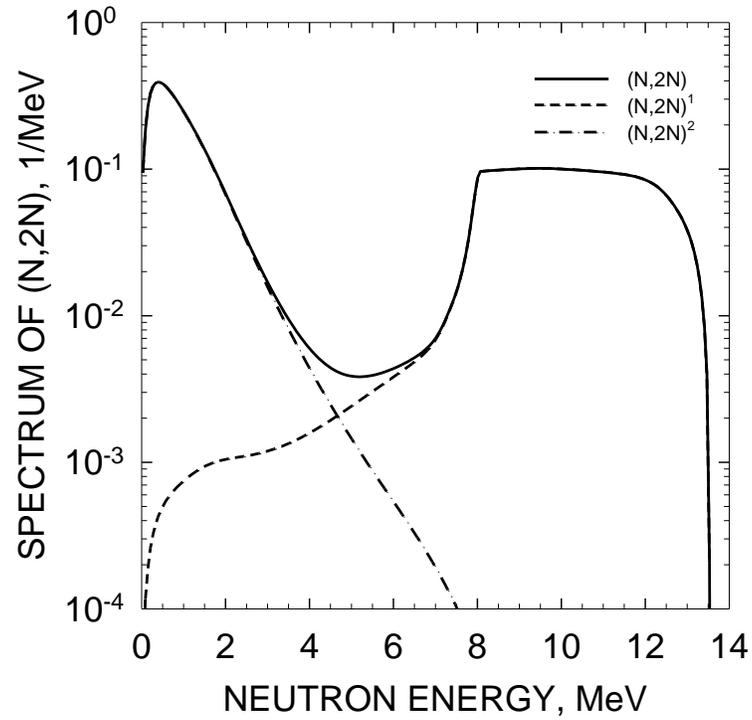

Fig. 12.10 (n, 2n) reaction neutron spectra of $^{243}$Am +n for incident neutron energy 20 MeV.

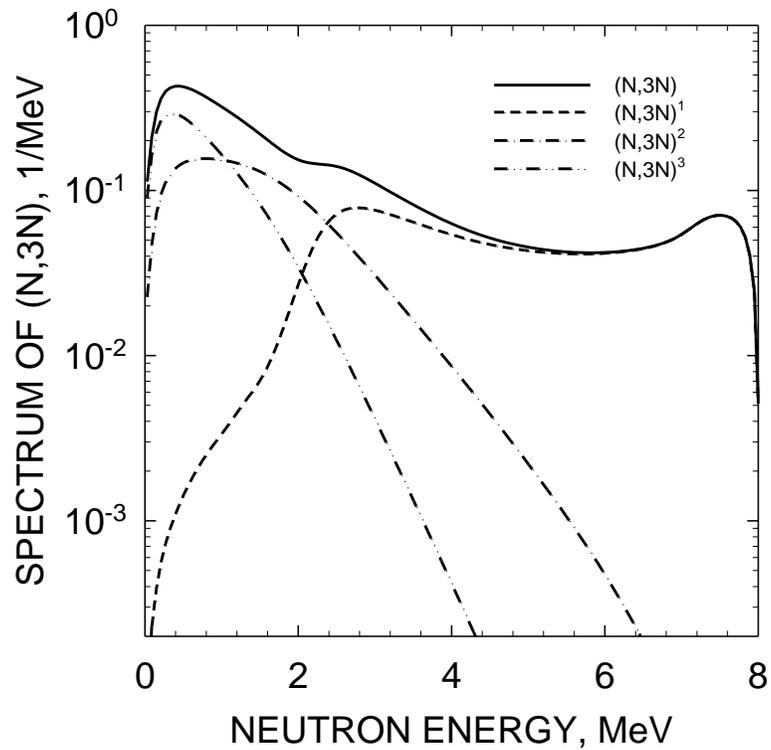

Fig. 12.11 (n, 3n) reaction neutron spectra of $^{243}$Am +n for incident neutron energy 20 MeV.

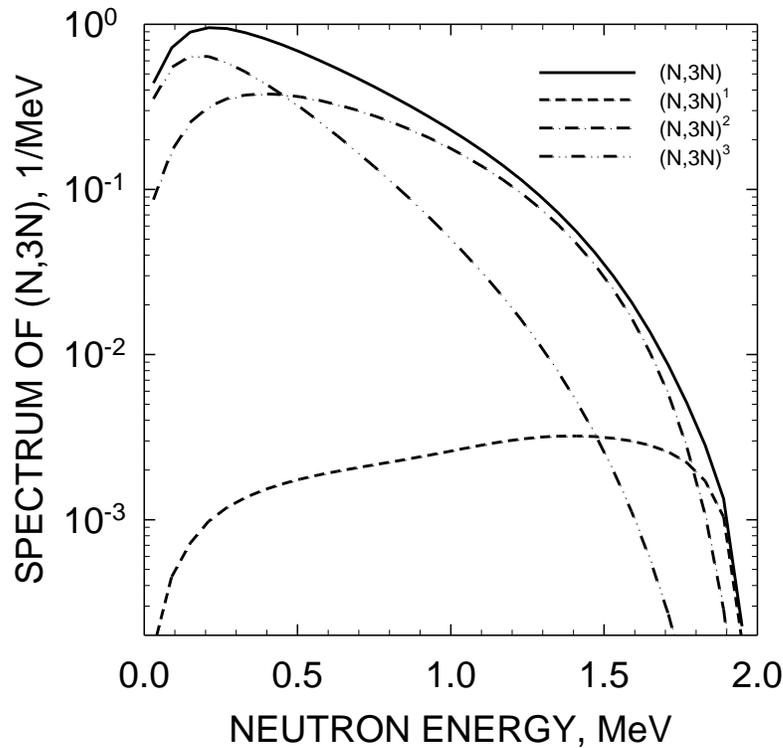

Fig. 12.12 Comparison of (n, 3n) reaction neutron spectra for $^{243}$Am +n
for incident neutron energy 14 MeV.

Spectrum of (n,nγ) reaction actually is just hard energy tail of 'pre-equilibrium' component of first neutron spectrum (see Figs. 12.1, 12.2, 12.3).

Spectrum of the first neutron of (n,2n) reaction is much softer, although 'pre-equilibrium' component still comprise appreciable part of it. Figures 12.10, 12.11 and 12.12 illustrate the variation of the partial contributions of the 1st and 2nd neutrons to the combined spectrum of $^{243}$Am (n,2n) and (n,3n) reactions. First neutron spectrum of (n,3n) reaction is actually of evaporative nature. First neutron spectrum of (n,nf) reaction has rather long pre-equilibrium high-energy tail. First neutron spectrum of (n,2nf) reaction, as that of (n,3n) reaction, is of evaporative nature. Figures 12.11 ($E_n$= 20 MeV) and 12.12 ($E_n$= 14 MeV) illustrate the variation of the partial contributions of the 1st, 2nd and 3d neutrons to the combined spectrum of $^{241}$Am (n,3n) reaction, softening of higher multiplicity neutrons is evident.

## 13. Conclusion

The diverse measured database of n+$^{243}$Am is analyzed using a statistical theory and generalized least squares codes. Important constraints for the measured capture cross section comes from the average radiative, $S_0$ and $S_1$ strength functions, however, the observed capture cross section needs anomalously high capture width to produce a consistency with measured data, fited with GMA code approach in keV-energy range.

Evaluated are the $^{243}$Am neutron-induced fission, capture, inelastic scattering, (n,2n) cross sections, branching ratio for the yields of short-lived $^{244m}$Am and long-lived $^{244g}$Am states of $^{243}$Am(n, γ) reaction, branching ratio for the yields of short-lived $^{242g}$Am and long-lived $^{242m}$Am states of $^{243}$Am(n, 2n) reaction. Prompt fission neutron spectrum (PFNS) and prompt fission neutron multiplicity are another important items predicted, measurements of the former for the [243]Am(n, F) reaction are unavailable at the moment.

Prompt fission neutron spectra data for the first-chance fission and emissive fission reactions are predicted. The influence of exclusive (n, xnf) pre-fission neutrons on prompt fission neutron spectra (PFNS) and (n, xn) spectra is modelled. Contributions of emissive/non-emissive fission and exclusive spectra of (n, xnf) reactions are defined by a consistent description of the [243]Am (n, F), [241]Am (n, F), [241]Am (n, 2n) [241]Am reactions. In [243]Am neutron capture the branching ratio data of the yields of short-lived ($1^-$) and long-lived ($6^-$) [244]Am states measured at thermal energy is calculated without arbitrary normalizations, though in a simplified manner. Excited levels of [244]Am are modelled using predicted Gallher-Moshkowski doublets. In [243]Am(n, 2n) reaction the branching ratio data of the yields of short-lived ($1^-$) and long-lived ($6^-$) [242]Am states are calculated without arbitrary normalizations, though in a simplified manner. Excited levels of [242]Am are modelled using predicted Gallher-Moshkowski doublets.

The job supported by the International Science and Technology Center (Moscow) under Project Agreement B-1604.Data file is at https://www-nds.iaea.org/minskact/data/original/za095243

## REFERENCES

1. Marie F., Letourneau A., Fioni G. et al. Nucl. Instr. and Meth. in Physics Research. A 556, pp. 547-555.
2. Bringer O., AlMahamid I., Chabod S., et al. Proc. of Int. Conf. on Nuclear Data for Science and Technology. Nice, France, Vol. 1, p. 619 (2008).
3. Maslov V.M., Porodzinskij Yu.V., Baba M., Hasegawa A. Nucl. Sci. Eng. 143, 177 (2003).
4. Maslov V. M. In: Proc. of the 13th International Seminar on Interaction of Neutrons with Nuclei, May 25-28, 2005, Dubna, Russia, p. 321.
5. Maslov V.M., Pronyaev V.G., Tetereva N.A. et al. INDC(BLR)-021, 2010.
6. Maslov V.M., Pronyaev V.M., Tetereva N.A. et al. In: Proc. Intern. Conf. on Nuclear Data for Science and Technology, Jeju, Korea, April 26-30, 2010, Journal of Korean Phys. Soc., **59**, 1337 (2011).
7. Maslov V.M., Pronyaev V.M., Tetereva N.A., Granier T., Hambsch F.-J. In: Proc. Intern. Conf. on Nuclear Data for Science and Technology, Jeju, Korea, April 26-30, 2010 Journal of Korean Phys. Soc., **59**, 867 (2011).
8. Maslov V.M., Sukhovitskij E.Sh., Porodzinskij Yu.V., Klepatskij A.B., Morogovskij G.B. INDC(BLR)-006, 1996.
9. Poenitz W.P., Aumeier S.E., The simultaneous evaluation of the standards and other cross sections of importance for technology, Report ANL/NDM-139 (1997).
10. International Evaluation of Neutron Cross-Section Standards, IAEA, Vienna, Austria, IAEA Publication STI/PUB/1291 (2006).
11. Weston L.W. and Todd J.H. Nucl. Sci. Eng., 91, 444 (1985).
12. Wisshak K., Kappeler F. Nucl. Sci. Eng. 85, 251 (1983).
13. Shibata K., Iwamoto O., Nakagawa T. . Nucl. Sci. Techn., 48 (2011) 1.
14. Seeger P.A. Fission cross sections from Pommard. LA-4420, 1970, p. 138.
15. Behrens J.W., Browne J.C. Nucl. Sci. Engng. 77, 444 (1981)
16. Butler D.K. and Sjoblom R.K. Phys. Rev. 124, 1129 (1961).
17. Terayama H., Karino Y., Manabe F., Yanagawa M., Kanda K., Hirakawa N. NEANDC(J)-122, 8609.

18. Knitter H.-H., Budtz-Jorgensen C. Nucl. Sci. Engng. 99, 1 (1988).
19. Manabe F., Kanda K., Iwasaki T. Terayama H., Karino Y., Baba M., Hirakawa N. Fourn. Fac. of Eng., Tohoku University Tech. Rep. 52, 97, 1988.
20. Aiche M. et al. Proc. Intern. Conf. on Nuclear Data for Science and Technology, p. 483, Nice, France, April 22-27, 2007, EDP Sciences.
21. Fomushkin E.F., Gutnikova E.K., Zamyatnin Yu.S. et al. Yadernaya Fyzika, 5 (5), 966 (1967).
22. Fursov B.I., Baranov E.Yu., Klemyshev M.P. et al. Sov. J. At. Energy 59, 899 (1985).
23. Fomushkin E.F., Novoselov G.F., Vinogradov Yu.I. et al. Yadernye Konstanty 57(3), 17 (1984).
24. A.A.Goverdovsky, A.K. Gordyushin, B.D. Kuz'minov, V.F.Mitrofanov, A.I.Sergachev, S.M.Solov'ev, T.E.Kuz'mina, Atom. Energ. 67, (1), 30, 1989
25. Golovnya V.YA, Bozhko V.P., Oleynik S.N., Shpakov V.P., Kalinin V.A. JINR-E3-419, 293, 1999.
26. A.V. Fomichev, V.N. Dushin, S.M. Soloviov, A.A. Fomichev, S.G. Mashnik. LA-UR-05-1533. 2005
27. Laptev A.B., Scherbakov O.A., Vorobyev A.S., Haight R.C., Carlson A.D. Proc. 4-th Intern. Conf. Fission and Properties of neutron-rich nuclei, Sanibel Island, USA, 11-17 November, 2007.
28. Ignatjuk A.V., Istekov K.K., Smirenkin G.N. Sov. J. Nucl. Phys. 29, 450 (1979). 1984.
29. V.M. Maslov, INDC(BLR)-013/L, 1998, IAEA, Vienna.
30. Chadwick M., M. Herman, P. Oblozinsky et al. Nuclear Data Sheets. **112**, 2887-2996 (2011).
31. Maslov V.M. In: Proc. 13$^{th}$ International Symposium on capture gamma rays and related topics, Koln, August 25-29, 2008, Germany, American Instituite of Physics, 2009, p. 579.
32. Firestone R.B., Table of Isotopes CD-ROM, Eighth Edition, Version 1.0, March 1996, S.Y.Frank Chu, CD-ROM Ed., V.S.Shirley, Ed., Wiley-Interscience, 1996
33. Strutinsky V.M. Nucl. Phys.A95 (1967) 420
34. Phillips T.W., Howe R.E. Nucl. Sci. Eng. 69, 375 (1979)
35. Moldauer P.A. Phys. Rev., C11, 426 (1975).
36. Tepel J.W., Hoffman H.M., Weidenmuller H.A. Phys. Lett. 49, 1 (1974).
37. Fu C. Nucl. Sci. Eng. 86 (1984) 344.
38. Maslov V.M. Yad. Fiz. 63 (2000) 214.
39. Maslov V.M. Actinide Neutron-Induced Fission up to 20 MeV, Nuclear Reaction Data and Nuclear Reactors, International Center for Theoretical Physics Lecture Notes, 14 March – 14 April, Trieste, Italy, No.1, pp. 231-268 (2001)
40. Ignatyuk A.V., Maslov V.M. Sov. J. Nucl.Phys. .54 , p. 392, 1991.
41. Myers W.O. and Swiatecky W.J. Ark. Fyzik. 36 (1967) 243.
42. Bjørnholm S. and Lynn J.E. Rev. Mod. Phys. 52 (1980) 725
43. Howard W.M. and Möller P. Atomic Data and Nuclear Data Tables 25 (1980) 219.
44. Seeger P.A., Hemmendinger A., Diven B.C. Nucl. Phys. A96, 605 9549 (1967).
45. Kornilov N.V., Kagalenko A.V., Baryba V.Ya. et al. Ann. Nucl. Energy, 27, 1643 (2000).
46. Street K. Jr, Ghiorso A., Seaborg G.T. Phys. Rev. 79, p. 530 (1950).
47. Stevens C.M., Studier M.H., Fields P.R., et al. Phys. Rev. 94, p. 974 (1954).
48. Harvey B.G., Robinson H.P., Thompson S.G., et al. Phys. Rev. 95, p. 581 (1954).
49. Butler J.P., Lounsbury M., Merritt J.S., Canadian J. of Physics 35, pp. 147-154 (1957).
50. Vandenbosch R., Fields P.R., Vandenbosch S.E., Metta D. J. Inorg. Nucl. Chem. 26, pp. 219-234 (1964).
51. Ice C.H., Production of the transplutonium elements at Savannah River, Report DP-MS-66-69, 1966.

52. Schuman R.P, Resonance Activation Integral Measurements, Report IN-1126, p.19, December 1967.
53. Bak M.A.., Krivochatskiy A.S., Petrzhak K.A., et al.  Atomnaya Energija (Sov.), Vol. 23, issue 4, pp. 316-319, October 1967 (in Russian).
54. Folger R.L., Smith J.A., Brown L.S., et al. Proc. of Conf. Washigton, Vol. 2, p. 1279 (1968).
55. Simpson O.D., Simpson F.B., Harvey J.A., et al. Nucl. Sci. Eng. 55, pp.273-279 (1974).
56. Gavrilov V.D., Goncharov V.A., Ivanenko V.V., et al. Atomnaya Energya, Vol. 41, issue 3, pp. 185-190, September, 1976.
57. Hatsukawa Y., Shinohara N., Kentaro H., Measurements of Neutron Cross Section of the $^{243}$Am(n,$\gamma$)$^{244}$Am, Report JAERI-Conf 98-003, pp. 221-224 1998.
58. Ohta M., Nakamura S., Harada H. et al, Nucl. Sci. Eng. 43, pp. 1441-1445 (2006).
59. Hori J., Oshima M, Harada H, et al. Report JAERI-C-2009-004, p. 123, April 2009.
60. Cricchio A., Ernstberger R., Koch L., Wellum R., Proc. of the Int. Conf. Nuclear Data for Science and Technology, 6-10 September 1982, Antwerp, Holland, D.Reidel Publishing Company, p.175.
61. Sood P.C., Singh R.N. Nucl. Phys. A373, 519 (1982.
62. Strutinsky V.M., Groshev L.V., Akimova M.K., Atomnaya Energiya, 38, 598 (1960).
63. Uhl M., Strohmaier B., A Computer Code for Particle induced Activation Cross Sections and Related QuantitiesIRK-76/01,  IRK, Vienna (1976).
64. Maslov V.M., Porodzinskij Yu. V., Baba M., Hasegawa A., Kornilov N. V., Kagalenko A. B., Tetereva N.A. Phys. Rev. C, 69, 034607, 2004.
65. T.L. Norris, A.J. Gancarz, D.W. Efurd, R.E. Perrin, G.W. Knobeloch, P.W. Oliver, G.F. Grisham, R.J.  Prestwood, I., Binder, G.W. Butler, W.R. Daniels, and D.W. Barr. 14-MeV (n,2n) Neutron Cross Sections of $^{241}$Am and $^{243}$Am. In: Irradiations at the Rotating Target Neutron Source-II, p. 69. UCID-19837-83.
66. Dovbenko A.G., Ivanov V.I., Kolesov V.E., Tolstikov V.A., Bulleten Tsentra po Jadernym Dannym 6, pp. 42-46, M: Atomizdat 1969.
67. Khokhlov Yu. A., Ivanin I.A., In'kov V.I., et al. In:. Proc. Int. Conf. Nuclear Data for Science and Technology, Gatlinburg, USA, May 9-13, 1994, p. 272, J.K. Dickens (Ed.), ANS, 1994.
68. Frehaut J. et al., Progress Report CEA-N-2396, 1984, p. 69.
69. Malinovskij V.V. VANT (ser. Yadernie constanty), 1987, No.2, 25.
70. Saleh H.H., Parish T.A., Shinohara N., Nucl. Sci. and Eng., 125, p.51 (1997).
71. Piksaikin V.M., Balakshev Yu.F., Isaev S.G., Kazakov L.E. et al., Atomnaya Energiya, 85, 51 (1998).
72. Kornilov N.V., Kagalenko A. B, Hambsch F.-J.,. Phys. At.   Nucl., 1999, 62, 209.
73. Maslov V.M., Kornilov N.V., Kagalenko A. B., Tetereva N.A. Nucl. Phys. A760 (2005) 274.
74. Watt B.E. Phys. Rev. 87, 1037 (1952).
75. Madland D., Nix J., Nucl. Sci. Engng. 81, 213 (1982).